\documentclass[fleqn,usenatbib]{mnras}

\usepackage{newtxtext,newtxmath}
\usepackage[T1]{fontenc}

\DeclareRobustCommand{\VAN}[3]{#2}
\let\VANthebibliography\thebibliography
\def\thebibliography{\DeclareRobustCommand{\VAN}[3]{##3}\VANthebibliography}

\usepackage{graphicx}	% Including figure files
\usepackage{amsmath}	% Advanced maths commands
\usepackage{xcolor} % ADDED FOR RED REFEREE-REVISION HIGHLIGHTING
\title[Magnus Mountains on Neutron Stars]{Magnus mountains on spinning neutron stars}

\author[Y. Gangwar and D.I. Jones]{
Yashaswi Gangwar \thanks{E-mail: yg2n21@soton.ac.uk}
and D.I. Jones \thanks{E-mail: d.i.jones@soton.ac.uk}
\\
Mathematical Sciences and STAG Research Centre, University of Southampton, Southampton SO17 1BJ, UK\\
}

\date{Accepted XXX. Received YYY; in original form ZZZ}

\pubyear{2023}

\begin{document}
\label{firstpage}
\pagerange{\pageref{firstpage}--\pageref{lastpage}}
\maketitle


\begin{abstract}

We investigate the formation of a ``Magnus mountain'' on a neutron star, arising from the non-axisymmetric Magnus force acting on the elastic crust by pinned superfluid vortices.  Such a deformed star would act as a source of continuous gravitational waves. We consider a compressible two-component stellar model and solve the coupled equations of motion for the fluid and elastic components self-consistently, allowing the stellar deformation and the resulting mass quadrupole to be determined. For simplicity, we model the star as an infinitely long cylinder, consisting of a fluid region with a thin ocean and crust. We find that the current quadrupole is zero, while the Magnus forces are strong enough to produce (dimensionless) mass quadrupoles as large as $\sim 10^{-5}$ \textcolor{black}{within our cylindrical model}, and so would be limited only by the finite strength of the vortex pinning and the breaking strain of the star's elastic crust. \textcolor{black}{This ellipticity estimate serves as a proof of concept for the formation of Magnus mountains in a two-fluid model, and motivates extending the calculation to more realistic stellar models with a spherical geometry.}

%This amplitude is potentially large enough to produce continuous gravitational-wave signals that could be detectable by current gravitational-wave detectors. However, this estimate should be interpreted with caution, as it is based on a cylindrical star model. Although a direct comparison with earlier spherical results is not possible, this calculation provides an explicit estimate of the mass multipole in a fully compressible treatment. This framework may also be relevant in the modelling of pulsar glitches, as it highlights how non-axisymmetric vortex pinning and unpinning can lead to asymmetric stresses in the crust, potentially contributing to glitch phenomena. 

\end{abstract}


%%%%%%%%%%%%%%%%%%%%%%%%%%%%%%%%%%%%%%%%%%%%%%%%%%

%%%%%%%%%%%%%%%%% BODY OF PAPER %%%%%%%%%%%%%%%%%%

\section{Introduction}

Neutron stars provide a unique laboratory to study matter under conditions inaccessible on Earth. In terrestrial laboratories, matter can be studied at low density and high temperature, whereas neutron stars allow direct investigation of matter at extremely high densities and relatively low temperatures \citep{lattimer2021}. Although their interior temperatures are of order $10^8$ K, neutron stars are still considered extremely cold objects due to the very high Fermi energy which is a consequence of their extreme density \citep{SHAPIRO, Baym1969, Haskell_2018}. 

It is widely accepted  that the neutron fluid inside a neutron star exists in a state of superfluidity \citep{Migdal1959, Pines1985}. Many models also suggest that the proton fluid in the core is in a state of superconductivity \citep{Baym1969, Chamel2008}. It has been confirmed in lab experiments that a rotating superfluid mimics bulk rotation by forming quantised vortices \citep{Yarmchuk1979}.  Similarly, protons in the inner core may exist in a type-II superconducting phase, which gives rise to quantised magnetic flux tubes due to the internal magnetic field \citep{Baym1969, Haskell_2018, chamel2017superfluidity}.

A widely accepted model for pulsar glitches involves sudden unpinning of superfluid vortices, which transfer angular momentum to the crust, leading to a rapid spin-up of the star \citep{Haskell2015}. The observation of glitches provide strong evidence for the presence of pinned superfluid vortices in neutron stars. However, it has been argued that pinning in the crust alone is insufficient to account for the magnitude of observed glitches \citep{Andersson_2012, nicolas}. This has led to the suggestion that vortex pinning may also occur in the core, where neutron vortices can interact with magnetic flux tubes associated with the superconducting proton component. The dynamics of this vortex--flux-tube interaction were investigated by \cite{Ruderman1998}, who discussed that pinning and creep of vortices against flux tubes can significantly modify the coupling between the superfluid and charged components, thereby playing an important role in the rotational dynamics and glitch behaviour of neutron stars.

In this paper, we focus on modelling the formation of quadrupolar deformations, commonly referred to as mountains, sourced by the Magnus force in neutron stars. This provides a mechanism that can generate continuous gravitational waves (CGWs), which may be detectable by current gravitational-wave detectors. 

In a spinning-down neutron star, quantised neutron vortices move outwards and may become pinned either to nuclei in the inner crust or to magnetic flux tubes in the superconducting core \citep{Sauls1989,Ruderman1998,Haskell2015,Andersson2006}. Under perfect pinning, the vortex lines corotate with the charged component, which is itself tightly coupled to the crust through electromagnetic interactions. As the star spins down, the neutron superfluid cannot relax by moving its vortices outwards, and a rotational lag develops between the neutron superfluid and the charged component. Consequently, the vortices experience a Magnus force proportional to the relative velocity between the neutron superfluid and the vortex lines (or equivalently the charged component under perfect pinning). This force plays a central role in standard models of pulsar glitches by driving vortex unpinning and the subsequent transfer of angular momentum to the crust.

The idea of a Magnus mountain was first proposed by \citet{Jones_2002}, who argued that if vortices are pinned in a non-axisymmetric manner, the resulting rotational lag gives rise to a non-axisymmetric Magnus force. The pinned vortices exert an equal and opposite reaction force on the elastic crust, producing a non-axisymmetric deformation, or ``Magnus mountain'', which may act as a source of continuous gravitational waves.

A closely related idea was taken up in \citet{Melatos_2015}, who performed quantum mechanical simulations of vortex motion in pulsar glitches, and found that non-axisymmetries naturally arose.  They computed the current quadrupole radiation from such asymmetries, finding that detection might be possible by third-generation gravitational wave detectors, but only if pulsars have very large glitches, or if large glitches occur in millisecond pulsars.

Closer to the original Magnus mountain idea of \citet{Jones_2002},  \cite{universe8120619} investigated the mass and current multipoles generated by the non-axisymmetric pinning of vortices to flux tubes in the outer core of a neutron star. They considered a spherical two-fluid star consisting of a neutron superfluid (the ``normal'' component) and a charged component (proton + electron fluid). In their model, the neutron superfluid velocity was prescribed by hand in terms of Heaviside step functions. They concluded that for an incompressible, uniform-density star, the mass multipole vanishes because $\delta\rho = 0$, while the current multipole vanishes owing to the assumed symmetry of the prescribed velocity field. To estimate the mass quadrupole, the authors subsequently obtain the pressure perturbation from the incompressible Euler equation and infer the density perturbation using the compressible relation $\delta\rho \approx \delta p/c_s^2$. As acknowledged by the authors, this provides only an order-of-magnitude upper-limit estimate, yielding an ellipticity of $\sim 10^{-9}$, while the current multipole remains zero. Since the pressure perturbation is obtained from an incompressible model whereas the density perturbation is inferred using a compressible relation, the procedure is not fully self-consistent.  In contrast, we consider an infinite cylindrical two-component compressible star and solve the coupled equations of motion to calculate both the elastic response and the resulting shape change of the star.

\textcolor{black}{A recent study by \citet{Giliberti2026} estimated Magnus-force-induced mountains in spherical neutron-star models using realistic SLy4 and BSk21 equations of state. The calculation adopts a one-component, single-displacement description of the stellar response and therefore differs from the two-fluid model considered in this paper. As this provides the closest existing comparison with our calculation, we discuss the differences between the resulting ellipticities later in the paper.}

In our model, we assume an infinitely long cylindrical star of finite radius R. This choice reflects the natural geometry of the vortices, which are aligned along the rotation axis. This simplifies the analysis and allows us to capture essential physics of vortex pinning and deformation while avoiding spherical complications at this stage.  

\begin{figure}
    \centering    
    \includegraphics[width=\columnwidth]{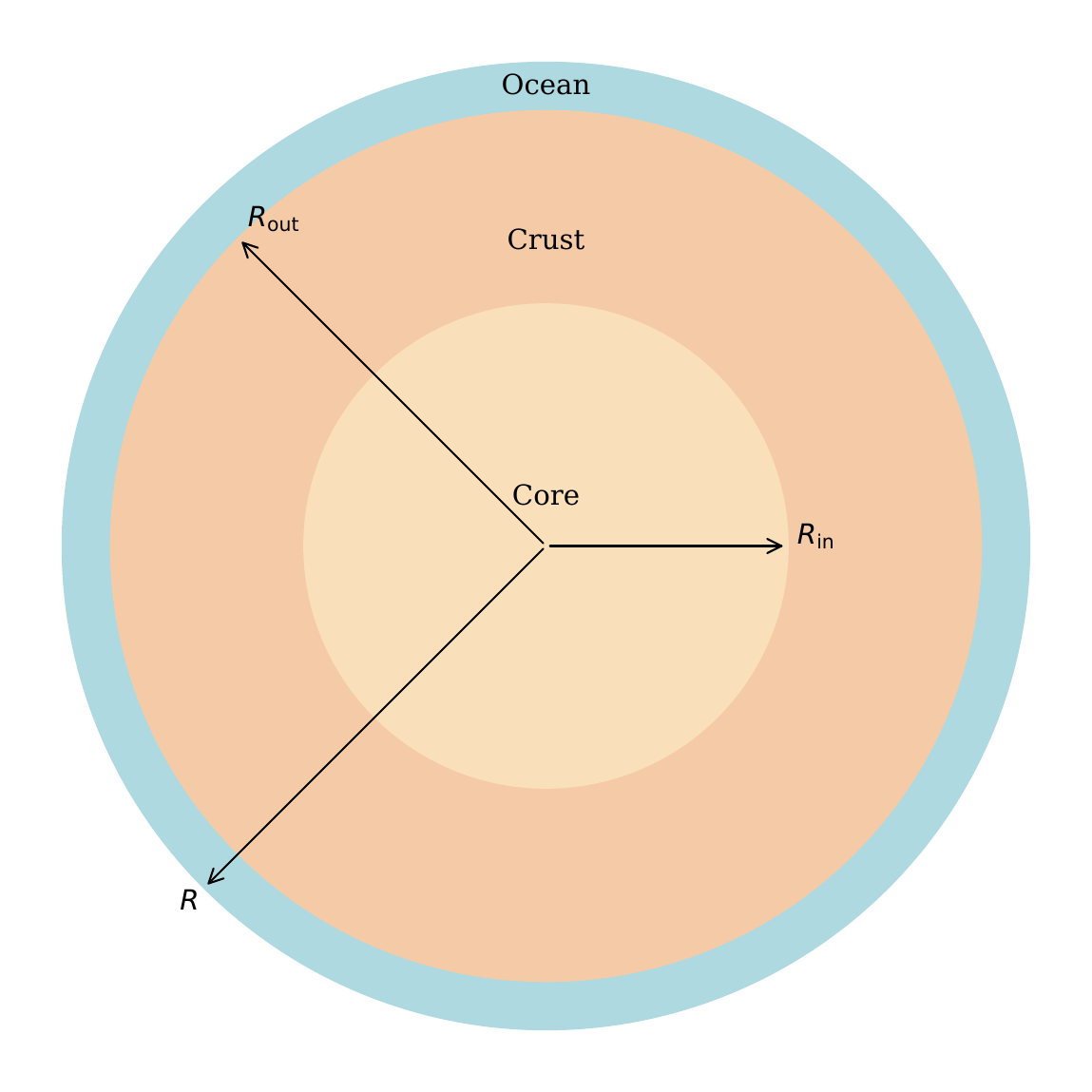}
    \caption{Schematic of the stellar model. Both the neutron superfluid and the charged component extend throughout the star. The charged component has a non-zero shear modulus only within the crust, occupying the annular region $R_{\mathrm{in}} < r < R_{\mathrm{out}}$. The regions $r \leq R_{\mathrm{in}}$ and $R_{\mathrm{out}} \leq r \leq R$ correspond to the fluid core and thin fluid ocean, respectively.}
    \label{fig:annular_prust}
\end{figure}

Our goal is to construct a simplified model for the formation of a Magnus mountain in a cylindrical star, with the intention of extending the analysis in future work to a more realistic spherical star. This paper focuses exclusively on the cylindrical case.

In Section~\ref{unper_eom}, we introduce the equation of state and present the unperturbed equations of motion that form the basis of our perturbative analysis. In section \ref{background star}, we describe the background star on which the perturbation will be imposed. In Section~\ref{sec8:pert_vel}, we introduce the superfluid velocity perturbation on top of the background star, which gives rise to a non-axisymmetric Magnus force acting on the crust through the pinned vortices.  We assume that the velocity perturbation of the charged component vanishes. In Section~\ref{Solving Elastic Equations of Motion}, we solve the coupled equations of motion (EOM) for the fluid and elastic components of the star. The two sets of equations are coupled through the Magnus force, resulting in a system of four coupled ordinary differential equations (ODEs). The boundary conditions used to solve this system are described in Section~\ref{bc}. 

In Section~\ref{quad}, we derive an expression for the quadrupole ellipticity, which is then used to estimate the size of the resulting mountain. In Section~\ref{BOTE}, we introduce the three dimensionless parameters that govern the model, and use these to make a back-of-the-envelope estimate for the crustal displacements induced by the Magnus force, which is later compared with the numerical results.
In Section~\ref{NR}, we present the numerical solutions and the corresponding displacement-field plots obtained by solving the system of coupled ordinary differential equations. We also present a fitting formula for the quadrupole ellipticity, and provide an upper limit on the size of the resulting mountain within our model. In Section ~\ref{crust_breaking}, we investigate whether the crust yields before vortex unpinning occurs by applying the von Mises criterion. Finally, in Section~\ref{summary_ponclusion}, we summarise the proposed model for the formation of Magnus mountains.

\section{Unperturbed Equations of motion}
\label{unper_eom}

\textcolor{black}{In our setup, we consider a cylindrical stellar configuration. We model the star as a two-component system: a neutron superfluid and a charged component consisting of protons and electrons. Both the neutron superfluid and the charged component extend throughout the star, occupying the region
\begin{equation}
\textcolor{black}{0 \leq \hat{r} \equiv \frac{r}{R} \leq 1,}
\end{equation}
\textcolor{black}{where $r$ is the dimensional radial coordinate and $R$ is the dimensional stellar radius. We use $\hat{r}$ to denote the dimensionless radial coordinate throughout the paper.} However, the charged component has a non-zero shear modulus only within the ``crust", occupying the annular region
\begin{equation}
\textcolor{black}{0.9 = \hat{R}_{\mathrm{in}} < \hat{r} < \hat{R}_{\mathrm{out}} = 0.99,}
\end{equation}
\textcolor{black}{Here $\hat{R}_{\mathrm{in}}=R_{\mathrm{in}}/R$ and $\hat{R}_{\mathrm{out}}=R_{\mathrm{out}}/R$.} Consequently, the region $r \leq R_{\mathrm{in}}$ is treated as a purely fluid core, while the region $R_{\mathrm{out}} \leq r \leq R$ represents a thin fluid ocean. This configuration is illustrated in Fig.~\ref{fig:annular_prust}. We assume that the vortices are pinned to the charged component in the crust, so that the Magnus force on the neutron superfluid is accompanied by an equal and opposite force on the charged component.} 

In our model, we adopt a simple two-fluid polytropic equation of state (EOS) where each fluid obeys an independent $n=1$ polytrope. The total internal energy density is given by:
\begin{equation}
\epsilon = \frac{1}{2}\sum_{X} K_X n_X^2,
\label{eq:energy_density}
\end{equation}
where $X =\mathrm{n}$ corresponds to the neutron component and $X =\mathrm{p}$ to the charged component. The quantities \(n_X\) and \(K_X\) represent the corresponding number density and polytropic constant, respectively. We neglect entrainment between the neutron and charged components. 

The chemical potentials for each component are obtained by differentiating the energy density with respect to the corresponding number density:
\begin{align}
\mu_X &=  K_X n_X,
\label{eq:mu_n}
\end{align}
We rewrite this relation in terms of the mass density $\rho_X = m_X n_X$, yielding:
\begin{align}
\mu_X &= \frac{K_X}{m_X} \rho_X .
\label{eq:mu_n_rho}
\end{align}
We define reduced chemical potential as
\begin{align}
\tilde{\mu}_X = \frac{\mu_X}{m_B},
\label{eq:reduced_mu_p}
\end{align}
where $m_B$ is the common baryon mass. Using \eqref{eq:mu_n_rho}, \eqref{eq:reduced_mu_p} becomes
\begin{align}
\Tilde{\mu}_X &= \frac{K_X}{m_X^2} \rho_X .
\label{eq:mu_n_tilde} 
\end{align}
We can invert this to obtain the mass density as:
\begin{align}
\rho_X &= \frac{m_X^2}{K_X} \Tilde{\mu}_X = a_X \Tilde{\mu}_X.
\label{eq:rho_n}
\end{align}
The total mass density is then
\begin{equation}
\rho = \sum_{X=n,p} a_X \tilde{\mu}_X.
\label{eq:rho_total}
\end{equation}

We now introduce the Euler equations governing the neutron and charged components in the rotating frame of the star \citep{Grosart2005, Andersson2008, Passamonti2012}. These equations describe the nonlinear dynamics of the two-fluid system. For the neutron superfluid component, the Euler equation reads  
\begin{equation}
 \partial_t \vec{v}_{\mathrm{n}}+ \vec{v}_{\mathrm{n}}\cdot \nabla \vec{v}_{\mathrm{n}}+ 2 \vec{\Omega} \times \vec{v}_{\mathrm{n}}+ \vec{\Omega} \times (\vec{\Omega} \times \vec{r})  = - \nabla \tilde{\mu}_{\mathrm{n}}- \nabla \Phi + \frac{\vec{F}^{\mathrm{Mag}}}{\rho_\mathrm{n}},
\label{eq:euler_rotating_1}
\end{equation}
and for the charged component we have  
\begin{equation}
 \partial_t \vec{v}_{\mathrm{p}}+ \vec{v}_{\mathrm{p}}\cdot \nabla \vec{v}_{\mathrm{p}}+ 2 \vec{\Omega} \times \vec{v}_{\mathrm{p}}+ \vec{\Omega} \times (\vec{\Omega} \times \vec{r})  =  \frac{\nabla\cdot\boldsymbol{\tau}}{\textcolor{black}{\rho_\mathrm{p}}} - \nabla \Phi - \frac{\vec{F}^{\mathrm{Mag}}}{\rho_\mathrm{p}}.
\label{eq:euler_rotating_2}
\end{equation}
Here, $\vec{v}_\mathrm{p}$ and $\vec{v}_\mathrm{n}$ denote the velocity fields of the charged fluid and the neutron superfluid, respectively, as measured in the rotating frame. The gravitational potential is denoted by $\Phi$, while $\vec{\Omega}$ represents the angular velocity vector of the star, and
$\boldsymbol{\tau}$ is the elastic stress tensor.

\textcolor{black}{
We write the inertial-frame velocities as}
\begin{equation}
\textcolor{black}{
\vec{u}_X=\vec{v}_X+\vec{\Omega}\times\vec{r},\qquad X=\mathrm{n},\mathrm{p}.}
\label{eq:inertial_velocity_def}
\end{equation}

\textcolor{black}{
It then follows that the neutron-fluid vorticity in the inertial frame is}
\begin{equation}
\textcolor{black}{
\nabla\times\vec{u}_{\mathrm n}=2\vec{\Omega}+ \nabla\times\vec{v}_{\mathrm n}.}
\label{eq:inertial_vorticity}
\end{equation}

\textcolor{black}{
The relative velocity between the neutron and charged components is unchanged between the two frames,}
\begin{equation}
\textcolor{black}{
\vec{u}_{\mathrm n}-\vec{u}_{\mathrm p}=\vec{v}_{\mathrm n} -
\vec{v}_{\mathrm p}.}
\label{eq:relative_velocity_frames}
\end{equation}
 The quantity
\begin{equation}
\label{eq:magnus_force}
    \vec{F}^{\rm Mag} = \rho_{\mathrm{n}}\left( \textcolor{black}{2\vec{\Omega}+}\nabla \times \vec{v}_{\mathrm{n}}\right) \times \left( \vec{v}_{\mathrm{n}}- \vec{v}_{\mathrm{p}}\right)
\end{equation}
represents the Magnus force acting on the neutron superfluid due to the presence of pinned vortices. 

Assuming that the background star is elastically relaxed, the stress tensor satisfies:
\begin{equation}
\label{tau_rel_incom}
\nabla\cdot\boldsymbol{\tau}= -\rho_{\mathrm{p}}\nabla\tilde{\mu}_{\mathrm{p}}.
\end{equation}

Using the density-chemical potential relation given by
equation~\eqref{eq:rho_n}, the above equation can be integrated to obtain the background stress tensor,
\begin{equation}
\label{stress_back_pomp}
\boldsymbol{\tau}=-\frac{a_{\mathrm{p}}\tilde{\mu}_{\mathrm{p}}^{2}}{2}\,\mathbf{I},
\end{equation}
where $\mathbf{I}$ is the identity tensor. Let us define,
\begin{equation}
\label{psi_pomp}
\Psi = \frac{a_{\mathrm{p}}\tilde{\mu}_{\mathrm{p}}^{2}}{2},
\end{equation}
the stress tensor can be written more compactly as
\begin{equation}
\boldsymbol{\tau}=-\Psi\,\mathbf{I}.
\end{equation}

The gravitational potential satisfies Poisson's equation,
\begin{equation}
\nabla^2 \Phi = 4 \pi G \rho .
\label{eq:poisson_A1}
\end{equation}

The equations introduced in this section form the basis of the analysis presented in the remainder of the paper. In the next section, we solve them to determine the background stellar configuration. We subsequently perturb the equations of motion about the background equilibrium and solve for the displacement field generated by the asymmetric Magnus force.

\section{Background star model}
\label{background star}

We now construct the background configuration for a compressible star, which forms the equilibrium state upon which the perturbations will be introduced. We choose the background configuration to be elastically relaxed, meaning that there are no pre-existing elastic stresses in the background configuration. The system rotates uniformly with angular velocity \( \Omega \) about the \( z \)-axis, and due to the assumed cylindrical symmetry, all background quantities depend only on the radial coordinate \( r \). Within this setup, we can solve for the chemical potentials \( \mu_\mathrm{n}(r) \) and \( \mu_\mathrm{p}(r) \), as well as the gravitational potential \( \Phi(r) \), which together describe the equilibrium structure of the background star.

In the background configuration, the fluids are assumed to be co-rotating, so \( \vec{v}_{\mathrm{n}}= \vec{v}_{\mathrm{p}}\). In the rotating frame, this implies \( \vec{v}_{\mathrm{n}}= \vec{v}_{\mathrm{p}}= 0 \), which immediately implies $\vec{F}^{\mathrm{Mag}} = 0$. In this case, equations~\eqref{eq:euler_rotating_1} and~\eqref{eq:euler_rotating_2} reduce to the same form, which can be written in the unified notation  

\begin{equation}
\nabla  \tilde{\mu}_X = -  \nabla \Phi + \nabla \left( \frac{1}{2} |\vec{\Omega} \times \vec{r}|^2 \right).
\label{eq:euler}
\end{equation}
Equation~\eqref{eq:euler} represents the balance between the pressure gradient, gravitational force, and centrifugal force in the rotating frame. Integrating ~\eqref{eq:euler} gives
\begin{align}
\Tilde{\mu}_{\mathrm{n}}+ \Phi - \frac{1}{2} \Omega^2 r^2 &= C_\mathrm{n},
\label{eq:euler_mu_n}\\
\Tilde{\mu}_{\mathrm{p}}+ \Phi - \frac{1}{2} \Omega^2 r^2 &= C_\mathrm{p}.
\label{eq:euler_mu_p}
\end{align}

Since both fluids share a common surface, we impose that $\Tilde{\mu}_{\mathrm{n}}= \Tilde{\mu}_{\mathrm{p}}= 0$ at the surface, yielding $C_{\mathrm{n}}= C_{\mathrm{p}}= C_0$. Therefore, throughout the interior of the star, both fluids share a common reduced chemical potential:
\begin{equation}
\Tilde{\mu}_{\mathrm{n}}= \Tilde{\mu}_{\mathrm{p}}\equiv \Tilde{\mu}_0.
\label{eq:mu_pommon}
\end{equation}
Thus, equation~\eqref{eq:euler_mu_n} becomes:
\begin{equation}
\Tilde{\mu}_0 + \Phi - \frac{1}{2} \Omega^2 r^2 = C_0.
\label{eq:fluid_eos}
\end{equation}
Substituting this relation into Poisson's equation,
equation~\eqref{eq:poisson_A1}, yields
\begin{equation}
\nabla^2 \Phi  = 4 \pi G (a_{\mathrm{n}}+a_\mathrm{p}) \Tilde{\mu}_0.
\label{eq:poisson_A}
\end{equation}
We define
\begin{equation}
\label{A_definition}
    A = 4 \pi G (a_{\mathrm{n}}+ a_\mathrm{p}),
\end{equation}
which allows us to rewrite equation~\eqref{eq:poisson_A} as
\begin{equation}
  \nabla^2 \Phi = A \Tilde{\mu}_0.
\label{eq:poisson_A_n}  
\end{equation}
Using equation~\eqref{eq:fluid_eos} in equation~\eqref{eq:poisson_A_n}, we obtain:
\begin{equation}
\nabla^2 \Tilde{\mu}_0 + A \Tilde{\mu}_0 = 2 \Omega^2.
\label{eq:mu_poisson}
\end{equation}
In cylindrical coordinates, this becomes:
\begin{equation}
\frac{d^2 \Tilde{\mu}_0}{dr^2} + \frac{1}{r} \frac{d \Tilde{\mu}_0}{dr} + A \Tilde{\mu}_0 = 2 \Omega^2.
\label{eq:mu_poisson_pyl}
\end{equation}

This is a non-homogeneous Bessel differential equation. The homogeneous solution is given by:
\begin{equation}
\Tilde{\mu}_0^{(h)}(r) = C_1 J_0 (\sqrt{A} r),
\label{eq:mu_homo}
\end{equation}
where $J_0$ is the Bessel function of the first kind of order zero.

\textcolor{black}{The general homogeneous solution also contains a term proportional to $Y_0(\sqrt{A}r)$, the Bessel function of the second kind of order zero. Near the centre, this behaves as}
\begin{equation}
\textcolor{black}{Y_0(\sqrt{A}r)=\frac{2}{\pi}\left[\ln\left(\frac{\sqrt{A}r}{2}\right)+\gamma\right]+O(r^2|\ln r|),\qquad r\to 0^+,}
\end{equation}
\textcolor{black}{where $\gamma$ is the Euler--Mascheroni constant. Since this expression diverges as $r\to0^+$, we set the coefficient of $Y_0$ to zero so that the chemical potential remains finite at the centre.}
 For the particular solution, we assume $\Tilde{\mu}_0^{(p)} = B$, leading to:
\begin{equation}
B = \frac{2 \Omega^2}{A}.
\label{eq:mu_particular}
\end{equation}
Thus, the full solution for $\Tilde{\mu}_0$ is:
\begin{equation}
\Tilde{\mu}_0(r) = C_1 J_0(\sqrt{A} r) + \frac{2 \Omega^2}{A}.
\label{eq:mu_total}
\end{equation}

From equation~\eqref{eq:fluid_eos}, the gravitational potential becomes:
\begin{equation}
\Phi(r) = \frac{1}{2} \Omega^2 r^2 - C_1 J_0(\sqrt{A} r) - \frac{2 \Omega^2}{A} + C_0.
\label{eq:Phi_general}
\end{equation}

\textcolor{black}{The exterior gravitational potential is }
\begin{equation}
\textcolor{black}{\Phi_{\mathrm{ext}}(r)=2G\lambda\ln\!\left(\frac{r}{R}\right),\qquad r\geq R,}
\end{equation}
 \textcolor{black}{ where $\lambda$ denotes the mass per unit length of the star. For the cylindrical model, it is obtained by integrating the total mass density
$\rho=\rho_{\mathrm n}+\rho_{\mathrm p}$ over the stellar cross-section,}
\begin{equation}
\textcolor{black}{
\lambda = 2\pi\int_0^R \rho(r)\,r\,dr.}
\label{eq:lambda_def}
\end{equation}
\textcolor{black}{Continuity of the gravitational potential at $r=R$, together with the surface condition $\tilde{\mu}_0(R)=0$ give the value of $C_0$ as}

\begin{equation}
C_0 = -\frac{1}{2} \Omega^2 R^2.
\label{eq:C0}
\end{equation}

Another boundary condition applied at the surface is the continuity of the derivative of the gravitational potential at \( r = R \), which gives:

\begin{equation}
C_1 = \frac{2 G \lambda - \Omega^2 R^2}{R \sqrt{A} J_1(\sqrt{A} R)},
\label{eq:C1_final}
\end{equation}

Substituting equation~\eqref{eq:C1_final}  back into equation~\eqref{eq:mu_total} yields the final expression for the reduced chemical potential:
\begin{equation}
\Tilde{\mu}_0(r) = \frac{2 G \lambda - \Omega^2 R^2}{R \sqrt{A} J_1(\sqrt{A} R)} J_0(\sqrt{A} r) + \frac{2 \Omega^2}{A}.
\label{eq:mu_final}
\end{equation}

The corresponding internal gravitational potential becomes:
\begin{equation}
\Phi(r) = \frac{1}{2} \Omega^2 (r^2 - R^2) - \frac{2 G \lambda - \Omega^2 R^2}{R \sqrt{A} J_1(\sqrt{A} R)} J_0(\sqrt{A} r) - \frac{2 \Omega^2}{A}.
\label{eq:Phi_final}
\end{equation}

Using the relation from equation~\eqref{eq:rho_n}, the mass density of each component can be written as
\begin{equation}
\label{comp_rho_n}
    \rho_X(r) = \frac{m_X^2}{K_X} \left( \frac{2 G \lambda - \Omega^2 R^2}{R \sqrt{A} J_1(\sqrt{A} R)} J_0(\sqrt{A} r) + \frac{2 \Omega^2}{A}\right).
\end{equation}

\textcolor{black}{Using the dimensionless radial coordinate $\hat{r}$, the density of each component can be written as}
\begin{equation}
\label{comp_rho_n_dimless_1}
    \rho_X(\hat{r}) = \frac{m_X^2}{K_X} \left( \frac{2 G \lambda - \Omega^2 R^2}{R \sqrt{A} J_1(\sqrt{A} R)} J_0(\sqrt{A} R \hat{r}) + \frac{2 \Omega^2}{A} \right).
\end{equation}
Note that these solutions are exact and do not assume a slow-rotation approximation.

Throughout this paper we adopt a neutron-to-charged-component mass-per-unit-length ratio of \(9:1\), consistent with the neutron-rich composition expected in neutron-star interiors \citep{SHAPIRO}.
We assume a stellar radius of
$R = 10^{6}\,\mathrm{cm} = 10\,\mathrm{km},$
and a volume-averaged density of
$\bar{\rho} = 10^{15}\,\mathrm{g\,cm^{-3}},$
 representative of neutron-star matter for our cylindrical stellar model.  These choices determine the mass per unit length:
\[
\lambda
%= \pi \left(10^{6}\,\mathrm{cm}\right)^{2}
%\left(10^{15}\,\mathrm{g\,cm^{-3}}\right)
\simeq 3.1 \times 10^{27}\,\mathrm{g\,cm^{-1}}
\simeq 1.6\,M_{\odot}\,(10\,\mathrm{km})^{-1}.
\]
In Appendix~\ref{Back_env},  we provide further details of our background solution, including the values  of \(K_\mathrm{n}\) and \(K_\mathrm{p}\).

%are given in equations~\eqref{k_n_def} and~\eqref{k_p_def}, respectively. The derivation of these estimates is provided in Appendix~\ref{Back_env}, where we determine the parameters using the assumed neutron-to-proton mass-density ratio together with the background equilibrium condition that the chemical potential vanishes at the stellar surface. 

\begin{figure}
\large
    \centering
    \includegraphics[width=\columnwidth]{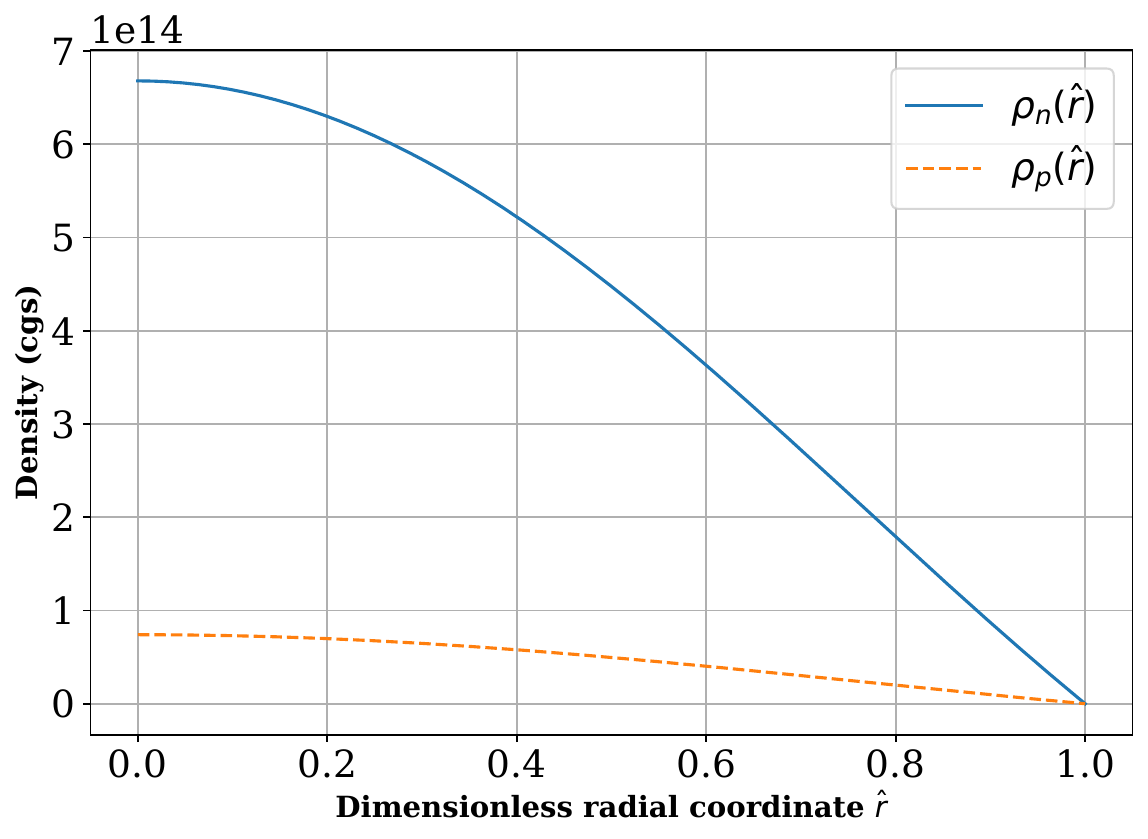}
   \caption{Plot of \( \rho_\mathrm{n}(\hat{r}) \) and \( \rho_\mathrm{p}(\hat{r}) \) as functions of the dimensionless radial coordinate $\hat{r}$. The angular velocity is taken to be \(\Omega = 100\,\mathrm{rad\,s^{-1}}\).  }

    \label{density_purve}
\end{figure}

The density profiles \(\rho_\mathrm{n}(\hat{r})\) and \(\rho_\mathrm{p}(\hat{r})\) are shown in Fig.~\ref{density_purve} as functions of the dimensionless radial coordinate \(\hat{r}\), for an
angular velocity of $\Omega = 100\,\mathrm{rad\,s^{-1}}.$  Both density profiles decrease monotonically with increasing \(\hat{r}\), from the center towards the stellar surface.

This completes the construction of the background equilibrium configuration of the star.
The resulting background quantities provide the foundation for the perturbative analysis presented in subsequent sections.  We will introduce linear perturbations about this equilibrium and solve the corresponding equations of motion to investigate the star’s response.

\section{The imposed m=2 superfluid velocity perturbation}
\label{sec8:pert_vel}

\begin{figure*}
    \centering
    \includegraphics[width=\linewidth]{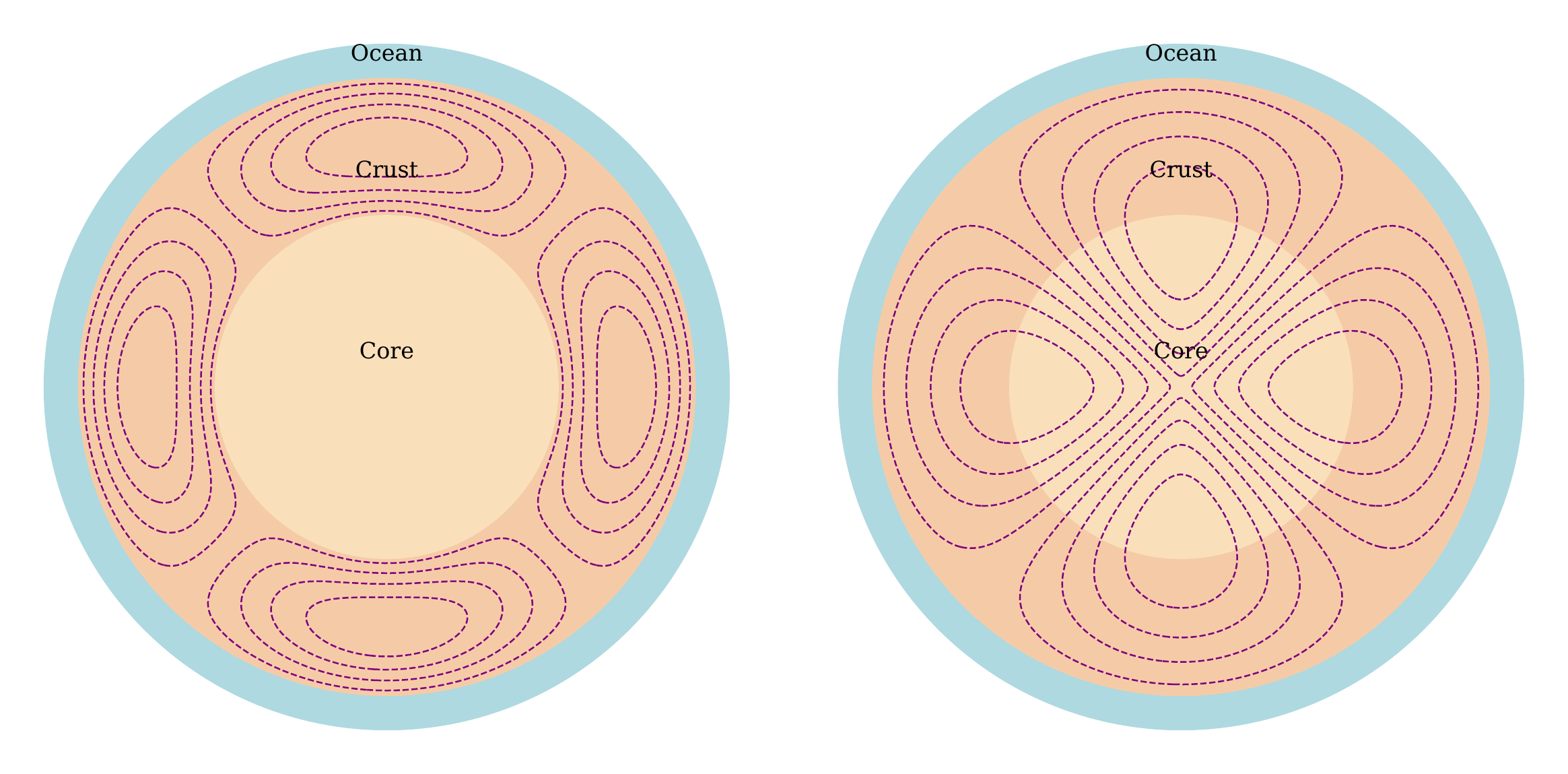}
    \caption{This figure shows two different cases of pinning regions. The left panel illustrates the case where pinning occurs only in the crust region, with the perturbation in the neutron velocity field confined accordingly. The right panel shows the scenario where pinning is present in both the crust and the core, and the corresponding velocity perturbation extends through both regions.}
    \label{pinn}
\end{figure*}

In this section, we introduce a non-axisymmetric perturbation to the neutron superfluid velocity field on top of the background rigid rotation. The perturbation is assumed to be confined to the crust, as illustrated in the left panel of Fig. \ref{pinn}. This perturbation is motivated by the non-axisymmetric pinning of vortices, which leads to a non-axisymmetric Magnus force on the charged component. \textcolor{black}{We seek a non-zero, stationary displacement of the
charged component in response to this force.  The resulting non-axisymmetric deformation can give rise to a ``mountain''.}

For the purposes of this study, we restrict our attention to azimuthal mode number \( m = 2 \), which is the natural cylindrical analogue of the quadrupolar \((l,m) = (2,2)\) perturbation in a rotating spherical star, where \(l\) and \(m\) are the spherical harmonic indices. \textcolor{black}{At the end of this section, we also evaluate the current quadrupole associated with the prescribed neutron-superfluid velocity perturbation.}

The neutron superfluid velocity perturbation field should ideally be calculated from first principles, taking into account the microphysical interactions of the vortices with the pinning potentials of flux tubes and lattice sites in the crust. One possible approach is to extend the model of asymmetric interactions between vortices and flux tubes presented by \cite{Sidery2009TheEO}, and to solve the system numerically by carefully analysing the vortex motion and distribution. This detailed calculation lies beyond the scope of this paper and is planned as part of future work. In this paper, we therefore specify the velocity field by hand.

To simplify the analysis, we ignore perturbations along the \( z \)-axis, restricting our attention to two-dimensional perturbations in the \( r \)--\( \phi \) plane. The perturbed velocity field is therefore given by:
\begin{equation}
    \vec{\delta v_\mathrm{n}}  = v_r(r) e^{i2\phi} \, \hat{r} + v_{\phi}(r) e^{i2\phi} \, \hat{\phi},
\label{eq:v_perturbed_general}
\end{equation}
where \( v_r \) and \( v_{\phi} \) are radial functions to be specified.

\textcolor{black}{
We seek a non-axisymmetric neutron-velocity perturbation whose fluid elements move along closed contours in the \(r\)--\(\phi\) plane. We introduce a scalar streamfunction \(\mathcal{S}(r,\phi)\), whose contours define the streamlines of the perturbed flow. A convenient form of the velocity perturbation is then}
\begin{equation}
\textcolor{black}{
\vec{\delta v}_{\mathrm{n}}
=
\frac{B_*}{\rho_\mathrm{n}(r)}
\,\hat{z}\times\nabla \mathcal{S},}
\label{eq:pert_v_def_final}
\end{equation}
\textcolor{black}{
where \(B_*\) is a constant. The factor \(1/\rho_\mathrm{n}(r)\) ensures that this velocity field satisfies the stationary neutron-fluid continuity equation in the rotating frame,}
\begin{equation}
\textcolor{black}{
\nabla\cdot\left(\rho_\mathrm{n}\vec{\delta v}_\mathrm{n}\right)=0.}
\label{eq:cont_pomp}
\end{equation}
\textcolor{black}{
The velocity field is tangent to the contours of constant \(\mathcal{S}\), so that these contours represent the streamlines of the perturbed neutron flow.}

We now specify the streamfunction $\textcolor{black}{\mathcal{S}}$ to ensure that the radial component of the perturbed velocity vanishes at the crust-core and crust-ocean boundaries, while keeping the azimuthal component finite. We adopt:
\begin{equation}
\label{psi_annulus}
    \textcolor{black}{\mathcal{S}} = \ \left( \frac{r^2}{R_{\text{in}}^2} -1 \right) \left( \frac{r^2}{R_{\text{out}}^2} -1 \right) \frac{r^2}{R^2} e^{2i\phi}.
\end{equation}
The leading-order behavior of equation~\eqref{psi_annulus} is proportional to \(r^2\) as \(r \to 0\), which ensures the regularity of the streamfunction at the origin.
We substitute equation~\eqref{psi_annulus} into equation~\eqref{eq:pert_v_def_final}, which yields the explicit form for the perturbed velocity components,

\begin{multline}
\label{per_vel_ann}
    \vec{\delta v_\mathrm{n}} = -\frac{2i B_*}{\rho_\mathrm{n}}\frac{r}{R^2}\left( \frac{r^2}{R_{\text{out}}^2} - 1\right) \left( \frac{r^2}{R_{\text{in}}^2} - 1\right) e^{2i\phi} \hat{r} 
    \\ + \frac{B_*}{\rho_\mathrm{n}}\left( \frac{6 r^5}{R^2 R_{\text{out}}^2 R_{\text{in}}^2}- \frac{4 r^3}{R^2 R_{\text{out}}^2}-\frac{4 r^3}{R^2 R_{\text{in}}^2} +  \frac{2r}{R^2} \right)e^{2i\phi} \hat{\phi}.
\end{multline}

For convenience, we introduce a compact notation by defining radial functions $\tilde{u}(r)$ and $\tilde{v}(r)$ as:
\begin{align}
\tilde{u}(r) &= \frac{2 B_*}{\rho_\mathrm{n}}\frac{r}{R^2}\left( \frac{r^2}{R_{\text{out}}^2} - 1\right) \left( \frac{r^2}{R_{\text{in}}^2} - 1\right),
\label{eq:uprime}\\
\tilde{v}(r) &= \frac{B_*}{\rho_\mathrm{n}}\left( \frac{6 r^5}{R^2 R_{\text{out}}^2 R_{\text{in}}^2}- \frac{4 r^3}{R^2 R_{\text{out}}^2}-\frac{4 r^3}{R^2 R_{\text{in}}^2} +  \frac{2r}{R^2} \right).
\label{eq:vprime}
\end{align}

Thus, equation~\eqref{per_vel_ann} can be rewritten compactly as:
\begin{equation}
\vec{\delta v}_{\mathrm{n}}= - i \tilde{u}(r) e^{i2\phi} \, \hat{r} + \tilde{v}(r) e^{i2\phi} \, \hat{\phi},
\label{eq:vn_pompact}
\end{equation}
where the factor of \(-i\) has been included for future convenience. 

\begin{figure}
    \centering
    \includegraphics[width=\columnwidth]{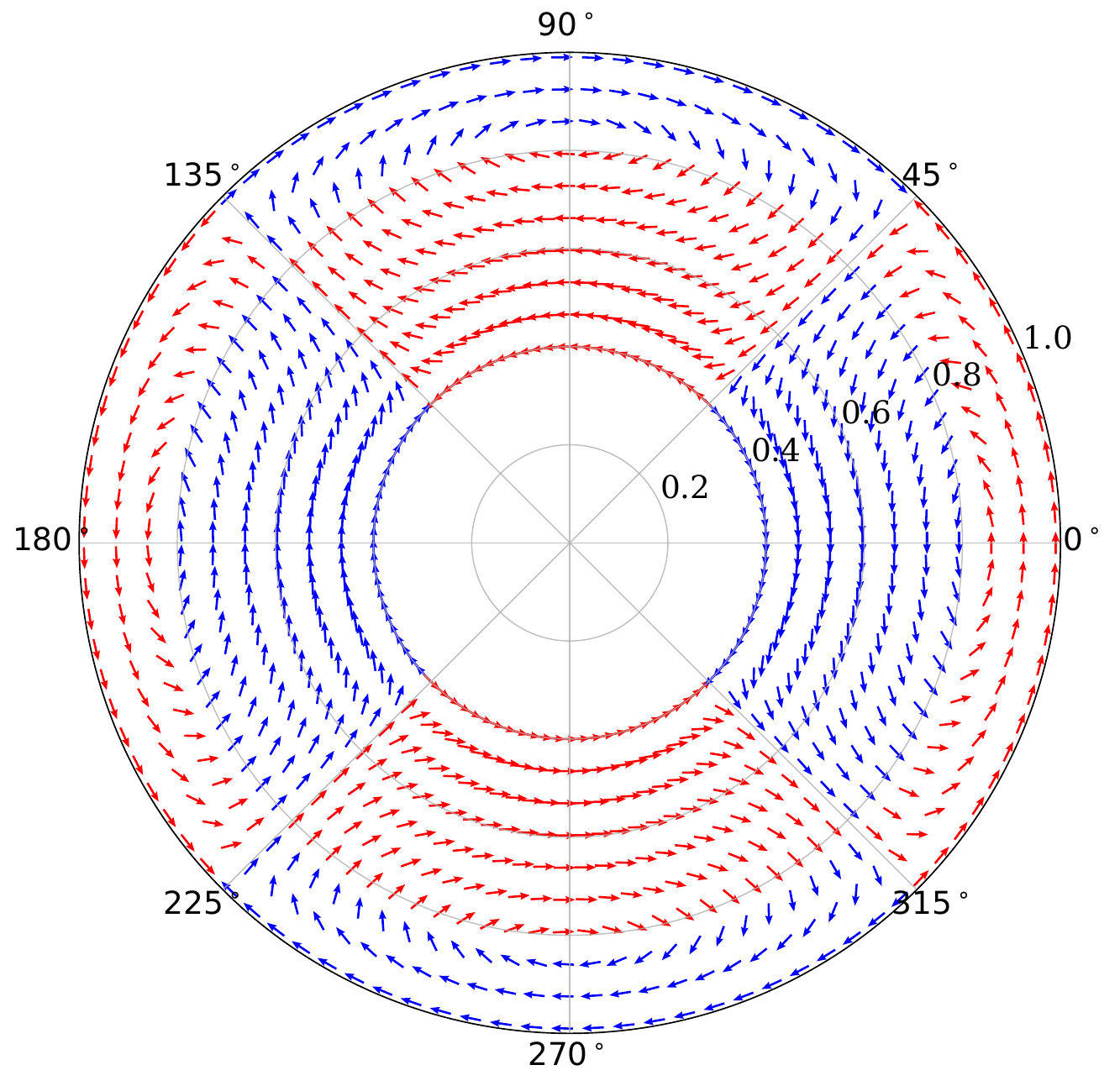}
    \caption{Perturbed superfluid neutron velocity field for the $m=2$ mode corresponding to equation~\eqref{per_vel_ann}. Arrows show the local direction of the velocity perturbation and have been normalised to equal length. Red and blue colours indicate positive and negative values of the azimuthal velocity perturbation, respectively. The crust thickness has been exaggerated for visual clarity and is not to scale.
}
    \label{per_vel_2}
\end{figure}

This gives an $m=2$ perturbed superfluid neutron velocity field appropriate for the compressible background configuration. Figure~\ref{per_vel_2} shows a visualisation of this perturbed velocity field. Note that, in Fig.~\ref{per_vel_2}, the crust thickness has been exaggerated in order to clearly visualise the flow of the velocity field. 

In the later sections, we will use this velocity perturbation to solve for the elastic displacement field that arises in response to this perturbation.

\textcolor{black}{At this stage, we cannot yet evaluate the mass quadrupole, since this requires solving the coupled perturbation equations to determine the resulting displacement field and density perturbation. However, the current quadrupole depends directly on the prescribed neutron-superfluid velocity perturbation and can therefore be evaluated at this point. The corresponding current multipole moment is defined by \citet{Thorne1980} as}
\begin{equation}
\textcolor{black}{
S_{22}
=
\frac{32\pi\sqrt{2}}{15}
\int
\rho_{\rm n}\,
\vec{\delta v}_{\rm n}\cdot
\vec{Y}^{\,*}_{B,22}\,
|\vec{x}|^2\,dV ,
}
\label{eq:s22_def}
\end{equation}
\textcolor{black}{where}
\begin{equation}
\textcolor{black}{
\vec{Y}^{\,*}_{B,22}
=
\frac{1}{\sqrt{6}}\,
\vec{x}\times\nabla Y^{*}_{22},
}
\end{equation}
\textcolor{black}{with}
\begin{equation}
\textcolor{black}{
Y_{22}
=
\frac{1}{4}
\sqrt{\frac{15}{2\pi}}
\sin^2\theta\,e^{2i\phi}.
}
\end{equation}

\textcolor{black}{For our cylindrical model, $|\vec{x}|^2=r^2+z^2$ and
$dV=r\,dr\,d\phi\,dz$. Since the velocity perturbation is confined to the crust, the current multipole becomes}
\begin{equation}
\textcolor{black}{
S_{22}
=
\frac{32\pi\sqrt{2}}{15}
\int_0^{z_{\rm max}}
\int_0^{2\pi}
\int_{R_{\rm in}}^{R_{\rm out}}
\rho_{\rm n}\,
\vec{\delta v}_{\rm n}\cdot
\vec{Y}^{\,*}_{B,22}\,
(r^2+z^2)\,
r\,dr\,d\phi\,dz .
}
\label{eq:s22_cyl}
\end{equation}

\textcolor{black}{Writing the magnetic-type vector spherical harmonic in cylindrical coordinates and using equation~\eqref{eq:vn_pompact}, we find}
\begin{equation}
\textcolor{black}{
\vec{\delta v}_{\rm n}\cdot\vec{Y}^{\,*}_{B,22}
=
\frac{1}{2\sqrt{6}}
\sqrt{\frac{15}{2\pi}}\,
\frac{zr}{r^2+z^2}
\left[\tilde{u}(r)+\tilde{v}(r)\right].
}
\label{eq:s22_dotproduct}
\end{equation}

\textcolor{black}{Substituting equation~\eqref{eq:s22_dotproduct} into equation~\eqref{eq:s22_cyl} and carrying out the $\phi$- and $z$-integrations gives}
\begin{equation}
\textcolor{black}{
S_{22}
=
\frac{16\pi\sqrt{\pi}}{\sqrt{90}}\,
z_{\rm max}^{\,2}
\int_{R_{\rm in}}^{R_{\rm out}}
\rho_{\rm n}
\left[
\tilde{u}(r)+\tilde{v}(r)
\right]
r^2\,dr .
}
\label{eq:s22_radial}
\end{equation}

\textcolor{black}{Using the compact velocity field together with the stationary neutron-fluid continuity equation~\eqref{eq:cont_pomp}, we obtain}
\begin{equation}
\textcolor{black}{
\tilde{v}(r)
=
\frac{1}{2\rho_{\rm n}}
\frac{d}{dr}
\left[
r\rho_{\rm n}\tilde{u}(r)
\right].
}
\label{eq:s22_v_u_relation}
\end{equation}

\textcolor{black}{Substituting this relation into equation~\eqref{eq:s22_radial} gives}
\begin{equation}
\textcolor{black}{
S_{22}
=
\frac{16\pi\sqrt{\pi}}{\sqrt{90}}\,
z_{\rm max}^{\,2}
\int_{R_{\rm in}}^{R_{\rm out}}
\left[
\rho_{\rm n}\tilde{u}(r)r^2
+
\frac{r^2}{2}
\frac{d}{dr}
\left(
r\rho_{\rm n}\tilde{u}(r)
\right)
\right]dr .
}
\end{equation}

\textcolor{black}{Integrating the second term by parts, the remaining volume terms cancel, leaving only the boundary contribution}
\begin{equation}
\textcolor{black}{
S_{22}
=
\frac{16\pi\sqrt{\pi}}{\sqrt{90}}\,
z_{\rm max}^{\,2}
\left[
\frac{1}{2}
\rho_{\rm n}\tilde{u}(r)r^3
\right]_{R_{\rm in}}^{R_{\rm out}} .
}
\end{equation}

\textcolor{black}{By construction, the radial component of the neutron-velocity perturbation vanishes at both boundaries of the pinning region, so that}
\begin{equation}
\textcolor{black}{
\tilde{u}(R_{\rm in})
=
\tilde{u}(R_{\rm out})
=
0 .
}
\end{equation}
\textcolor{black}{It therefore follows that}
\begin{equation}
\textcolor{black}{
S_{22}=0 .
}
\label{eq:s22_zero}
\end{equation}

\textcolor{black}{The current quadrupole $S_{22}$ vanishes because of the cylindrical
integration domain and the assumed neutron flow, which lies in the
$r$--$\phi$ plane and is independent of $z$. Mass conservation reduces
the radial integral to a boundary term. This term is zero because
the assumed streamfunction in equation~\eqref{psi_annulus} gives no radial neutron flow across
either boundary of the pinning region.} Thus, the $m=2$ neutron-velocity perturbation adopted in the present cylindrical model produces no current quadrupole.

\section{Assembling the perturbation equations}
\label{Solving Elastic Equations of Motion}

In this section, we assemble the coupled perturbation equations for the fluid and charged elastic components in the crust of the star. The elastic component in the crust is coupled to the fluid component through the Magnus force. The pinned vortices exert a reaction force on the elastic component that is equal in magnitude and opposite in direction to the Magnus force. \textcolor{black}{In response to this steady force, we expect
the charged component to develop a non-zero, stationary elastic
displacement, with $\delta\boldsymbol{v}_{\mathrm{p}}=0$ in the
corotating frame. }

We begin by perturbing the equation of motion for the neutrons, given in equation~\eqref{eq:euler_rotating_1}. The neutron superfluid velocity perturbation is prescribed explicitly, as defined in equation~\eqref{per_vel_ann}. We assume that the charge-fluid velocity perturbation vanishes, i.e. the (deformed) crust rotates rigidly.  We restrict our analysis to linear perturbations and adopt the Cowling approximation, whereby perturbations in the gravitational potential are neglected.  Under these assumptions, the perturbed Euler equation becomes:

\begin{equation}
2 \vec{\Omega} \times \vec{\delta v}_{\mathrm{n}}= -\nabla \delta \mu_{\mathrm{n}}+ \textcolor{black}{\frac{\delta \vec{F}^{\mathrm{Mag}}}{\rho_{\mathrm n}}},
\label{eq:perturbed_eom_rot}
\end{equation}
where $\delta\vec{F}^{\mathrm{Mag}}$ is the perturbed Magnus force.

We perturb the Magnus-force density given by
equation~\eqref{eq:magnus_force}. To first order, the perturbed Magnus
force is

\begin{align}
\delta\vec{F}^{\mathrm{Mag}}
={}&
\delta\rho_{\mathrm n}
\left(\textcolor{black}{2\vec{\Omega}+}\nabla\times\vec{v}_{\mathrm n}\right)
\times
\left(\vec{v}_{\mathrm n}-\vec{v}_{\mathrm p}\right)
\nonumber+
\rho_{\mathrm n}
\left(\nabla\times\delta\vec{v}_{\mathrm n}\right)
\times
\left(\vec{v}_{\mathrm n}-\vec{v}_{\mathrm p}\right)
\nonumber\\
&+
\rho_{\mathrm n}
\left(\textcolor{black}{2\vec{\Omega}+}\nabla\times\vec{v}_{\mathrm n}\right)
\times
\left(
\delta\vec{v}_{\mathrm n}
-
\delta\vec{v}_{\mathrm p}
\right).
\label{eq:perturbed_magnus_general}
\end{align}

For the co-rotating background considered here,
$\vec{v}_{\mathrm n}=\vec{v}_{\mathrm p}$.
By assumption, $\delta\vec{v}_{\mathrm p}=0$.  It follows that the first two terms in
equation~\eqref{eq:perturbed_magnus_general} vanish, leaving
\begin{equation}
\delta\vec{F}^{\mathrm{Mag}}= \rho_{\mathrm n} \left(\textcolor{black}{2\vec{\Omega}+}\nabla\times\vec{v}_{\mathrm n}\right)\times \delta\vec{v}_{\mathrm n}.
\label{permagforce1}
\end{equation}
\textcolor{black}{For the uniformly rotating background, we have $\vec{v}_{\mathrm n}=0$ in the rotating frame, which reduced equation~\eqref{permagforce1} to}
\begin{equation}
\delta\vec{F}^{\mathrm{Mag}}=2\rho_{\mathrm n}\, \vec{\Omega}\times\delta\vec{v}_{\mathrm n}.
\label{eq:magnus_linear}
\end{equation}

By substituting equation~\eqref{eq:magnus_linear} into the perturbed Euler equation~\eqref{eq:perturbed_eom_rot}, we obtain:
\begin{equation}
2 \vec{\Omega} \times \vec{\delta v}_{\mathrm{n}}= -\nabla \delta \mu_{\mathrm{n}}+ 2 \vec{\Omega} \times \vec{\delta v}_\mathrm{n}.
\label{eq:perturbed_eom_rot_final}
\end{equation}

We observe that the Coriolis force on the left-hand side of the equation exactly balances the Magnus force on the right-hand side. Canceling these terms on both sides of equation~\eqref{eq:perturbed_eom_rot_final} yields
\begin{equation}
\nabla \delta \mu_{\mathrm{n}}= 0 \quad \Rightarrow \quad \delta \mu_{\mathrm{n}}= \text{constant}.
\label{eq:delta_mu_ponstant}
\end{equation}

At the stellar surface, the Lagrangian perturbation condition for the chemical potential is: 
\begin{equation}
\Delta \mu_{\mathrm{n}}= \delta \mu_{\mathrm{n}}+ \textcolor{black}{\xi_{\mathrm n}^r(R,\phi)} \nabla_r \mu_{\mathrm{n}}= 0,
\label{eq:lagrangian_bc}
\end{equation}
which implies:
\begin{equation}
\text{constant} = - \textcolor{black}{\xi_{\mathrm n}^r(R,\phi)} \nabla_r \mu_\mathrm{n}.
\label{eq:fluid_bc}
\end{equation}
\textcolor{black}{Here $\xi_{\mathrm n}^r$ denotes the radial Lagrangian displacement of the neutron component.} If we assume a free surface perturbation of the form:
\begin{equation}
\textcolor{black}{\xi_{\mathrm n}^r(R,\phi)} = \textcolor{black}{f(R)} e^{2 i \phi},
\label{eq:xi_form}
\end{equation}
then the right-hand side of equation~\eqref{eq:fluid_bc} varies as \( \exp{(2i\phi)} \), while the left-hand side is constant. This condition can only be satisfied if:
\begin{equation}
\textcolor{black}{\xi_{\mathrm n}^r(R,\phi)} = 0.
\label{eq:xi_zero}
\end{equation}
Hence, we conclude:
\begin{equation}
\delta \mu_{\mathrm{n}}= 0.
\label{eq:delta_mu_zero_rot}
\end{equation}

\textcolor{black}{ In the corotating frame, the linear relation
$\delta\boldsymbol{v}_{\mathrm{n}}
= \partial_t\boldsymbol{\xi}_{\mathrm{n}}$ in the crust interior
implies that we have non-zero radial and azimuthal components of the neutron
displacement wherever the corresponding velocity
components are non-zero. Figure~\ref{per_vel_2} illustrates this: the velocity
arrows have both components within the pinning region and
are tangent to its boundaries, where
$\partial_t\xi_{\mathrm{n}}^r=0$.
At the stellar surface $r=R$, the free-surface condition gives
$\xi_{\mathrm{n}}^r(R,\phi)=0$, as shown in equation~\eqref{eq:xi_zero}.}

We find that the perturbed chemical potential \( \delta \mu_{\mathrm{n}}\) vanishes under the set of assumptions we have made.  The resulting solution is simple but not trivial. The Coriolis force associated with the prescribed velocity perturbation \( \delta \vec{v}_{\mathrm{n}}\) is exactly balanced by the Magnus force. There is simply no force associated with the gradient of a perturbed chemical potential.
\textcolor{black}{ The result $\delta\tilde\mu_{\mathrm n}=0$, together with the polytropic relation $\delta\rho_{\mathrm n}=a_{\mathrm n}\delta\tilde\mu_{\mathrm n}$, implies that $\delta\rho_{\mathrm n}=0$. Thus, the non-axisymmetric neutron flow does not produce a bulk Eulerian density perturbation at this order.}

Next, we solve the equations of motion (EOM) for the elastic component of the crust. We perturb equation~\eqref{eq:euler_rotating_2} under the Cowling approximation and solve the resulting equations of motion in the rotating frame. As mentioned earlier, we assume that there is no perturbation to the velocity of the charged component, i.e.
\[
\delta \vec{v}_{\mathrm{p}} = 0,
\]
so that the crust continues to rotate rigidly. The perturbed equation of motion then takes the form
\begin{equation}
\label{eq:eom_elastic_pomp}
    \nabla^a \delta \tau_{ab} - \frac{\delta \rho_\mathrm{p}}{\rho_\mathrm{p}} \nabla^a \tau_{ab} - \delta F^{\mathrm{Mag}}_b = 0,
\end{equation}

Substituting the expression for $\delta F^{\mathrm{Mag}}$ from equation~\eqref{eq:magnus_linear} and the background stress tensor from equation~\eqref{stress_back_pomp} into equation~\eqref{eq:eom_elastic_pomp} yields:
\begin{equation}
\label{eq:eom_elastic_pomp_2}
    \nabla^a \delta \tau_{ab} + \delta \rho_{\mathrm{p}}\nabla_b \tilde{\mu}_{\mathrm{p}} = \rho_{\mathrm{n}}2 (\Omega \times \vec{\delta v}_\mathrm{n})_b,
\end{equation}
which can equivalently be written in index notation as:
\begin{equation}
\label{eq:eom_elastic_pomp_3}
    \nabla^a \delta \tau_{ab} + \delta \rho_{\mathrm{p}} \nabla_b \tilde{\mu}_{\mathrm{p}}= \rho_{\mathrm{n}}2 \Omega \epsilon_{bzd} \delta v_\mathrm{n}^d.
\end{equation}

The perturbed stress tensor is given by
\begin{equation}
\label{eq:pert_stress_tensor_pom}
\delta \tau_{ab} = - \rho_{\mathrm{p}}\delta \tilde{\mu}_{\mathrm{p}}g_{ab} + \mu \left( \nabla_a \xi^{\textcolor{black}{\mathrm{p}}}_b + \nabla_b \xi^{\textcolor{black}{\mathrm{p}}}_a - \frac{2}{3} g_{ab} \nabla^c \xi^{\textcolor{black}{\mathrm p}}_{\textcolor{black}{\mathrm{c}}}\right),
\end{equation}
\textcolor{black}{In the elastic calculation, the Lagrangian displacement corresponds to the charged component, $\boldsymbol{\xi}^{\textcolor{black}{\mathrm p}}$. For simplicity, we write $\boldsymbol{\xi}^{\textcolor{black}{\mathrm p}} \equiv \boldsymbol{\xi}$, with $\xi_r$ and $\xi_\perp$ denoting its radial and tangential amplitudes.}
where the Lagrangian displacement can be written as:
\begin{equation}
\label{xi_definition}
\xi^a = \xi_r C_{m} \hat{r}^a + \xi_\perp r \nabla^a C_{m},
\end{equation}
with the harmonic factor defined as:
\begin{equation}
C_{m} = e^{im\phi},
\end{equation}
where  \(m\) is the azimuthal mode number. In this work, we focus on the case \(m=2\).

We adopt the expression for the shear modulus \( \mu(r) \) from equation~(13) of \cite{Strohmayer1991}, which is
\begin{equation}
\label{shear}
    \mu(r) = C^* \, \Tilde{\mu}_\mathrm{p}^{\frac{4}{3}}, 
\end{equation}
where the constant \( C^* \) is defined as
\begin{equation}
\label{shear_p}
C^* = 0.1194 \, e^2 \left( \frac{4\pi}{3} \right)^{\frac{1}{3}} 
\left( \frac{m_\mathrm{p}}{K_\mathrm{p}} \right)^{\frac{4}{3}}, 
\quad \text{(CGS units)}.
\end{equation}

%The electron charge $e$ is expressed in CGS units (statcoulomb). The reduced chemical potential $\tilde{\mu}_\mathrm{p}$ is also expressed in CGS units in equation~\eqref{shear}. 
In the original reference, the shear modulus was written in terms of the total density $\rho$. Since, in the two-fluid description adopted here, the elastic properties are associated only with the crustal (charged) component, we replace \(\rho\) by the crustal density \(\rho_\mathrm{p}\). We then rewrite the shear modulus, given in equation~\eqref{shear}, in terms of the reduced chemical potential \(\tilde{\mu}_\mathrm{p}\) using the relation between \(\rho_\mathrm{p}\) and \(\tilde{\mu}_\mathrm{p}\) given in equation~\eqref{eq:rho_n}. This modification provides a natural extension to the two-fluid framework. The expression for the shear modulus in equation~\eqref{shear} is applicable only within the annular crustal region, i.e., for \(R_{\mathrm{in}} \leq r \leq R_{\mathrm{out}}\), and vanishes outside this region.

\textcolor{black}{As a consistency check on this prescription, evaluating the shear modulus at the midpoint of the crust, $\hat r=0.95$, gives $\mu\simeq8\times10^{29}\,\mathrm{erg\,cm^{-3}}$. This is close to the typical crustal shear-modulus scale of $\sim10^{30}\,\mathrm{erg\,cm^{-3}}$ quoted in the literature \citep{Ushomirsky_2002}, suggesting that the use of the charged-component density provides a reasonable effective prescription within our two-fluid model.}

Inserting equations~\eqref{eq:pert_stress_tensor_pom} and \eqref{eq:vn_pompact} into equation~\eqref{eq:eom_elastic_pomp_3}, and employing the cylindrical-harmonic representation of the perturbed stress tensor developed in Appendix~\ref{app_pert_eom}, where the divergence of the stress tensor is evaluated and projected onto the radial and azimuthal directions, we obtain the radial and azimuthal components of the equations of motion:

\begin{align}
\frac{d\delta \tau_{rr}}{dr} - \frac{2\mu}{r}\left(\frac{\xi_r}{r} - \frac{d\xi_r}{dr}\right) - \frac{4}{r} \delta \tau'_{r\phi} + 8\mu \frac{\xi_\perp}{r^2} + \delta \rho_{\mathrm{p}}\frac{d \Tilde{\mu_\mathrm{p}}}{dr}
&= - \rho_{\mathrm{n}}2\Omega \tilde{v}(r), \label{eq:emo_elastic_r_2_real_pomp} \\
\delta \tau_{rr} + 2\mu\left(\frac{\xi_r}{r} - \frac{d\xi_r}{dr}\right) + r \frac{d \delta \tau'_{r\phi}}{dr} + 2 \delta \tau'_{r\phi} - 8\mu \frac{\xi_\perp}{r}
&= -r\rho_{\mathrm{n}} \Omega \tilde{u}(r). \label{eq:eom_elastic_phi_2_real_pomp}
\end{align}

We aim to solve the EOMs~\eqref{eq:emo_elastic_r_2_real_pomp} and \eqref{eq:eom_elastic_phi_2_real_pomp}. First, we substitute the expressions for the perturbed stress tensor components, equations~\eqref{tau_rr} and \eqref{tau_rperp}, into equation~\eqref{eq:emo_elastic_r_2_real_pomp}, obtaining:
\begin{multline}
\label{eq:elastic_eom_1}
    \frac{4 \mu}{3}\frac{d^2 \xi_r}{d r^2}  + \frac{4}{3}\frac{ d \xi_r}{  d r} \left( \frac{d \mu}{d r} + \frac{\mu}{r} \right) - \frac{2 \xi_r}{3 r} \left( \frac{d \mu}{d r} + \frac{8 \mu}{r} \right) + \frac{4 \xi_\perp}{ 3 r} \left( \frac{2\, d \mu}{d r} + \frac{7 \mu}{r} \right)\\ - \frac{4 \mu }{ 3 r } \frac{d \xi_{\perp}}{ dr} 
    - \rho_{\mathrm{p}}\frac{d \tilde{\delta \mu_\mathrm{p}}}{dr} - \tilde{\delta \mu_\mathrm{p}} \left( \frac{d \rho_\mathrm{p}}{d r} + D J_1(\sqrt{A} r )\right)  = -2 \rho_{\mathrm{n}}\Omega \tilde{v}(r).
\end{multline}

Similarly, substituting equations~\eqref{tau_rr} and \eqref{tau_rperp} into equation~\eqref{eq:eom_elastic_phi_2_real_pomp} yields:
\begin{multline}
\label{eq:elastic_eom_2}
    \mu \frac{d^2 \xi_\perp}{dr^2} + \frac{d \xi_{\perp}}{d r} \left( \frac{d \mu}{d r} + \frac{7 \mu}{3 r}\right) - \frac{\xi_{\perp}}{r} \left( \frac{d \mu}{d r} + \frac{19 \mu}{3 r}\right) + \frac{\mu}{3 r} \frac{d \xi_r}{d r} \\ + \frac{\xi_r}{r} \left( \frac{d \mu}{d r} + \frac{7 \mu}{3 r}\right) - \frac{\rho_{\mathrm{p}}\tilde{\delta \mu_\mathrm{p}}}{r} 
    = - \Omega \rho_{\mathrm{n}}\tilde{u}(r).
\end{multline}

In addition to the second-order EOMs~\eqref{eq:elastic_eom_1} and \eqref{eq:elastic_eom_2}, we also consider the continuity equation:
\begin{equation}
   \delta \rho_{\mathrm{p}}+  \nabla \cdot (\rho_{\mathrm{p}}\vec{\xi}) = 0,
\end{equation}
which simplifies to:
\begin{equation}
\label{xi_r_phi_2}
   \xi_{\perp} =  \frac{\delta \Tilde{\mu_\mathrm{p}} \, r}{4 \Tilde{\mu_\mathrm{p}}} + \frac{r \xi_r}{4 \rho_\mathrm{p}} \frac{d \rho_\mathrm{p}}{dr} + \frac{\xi_r}{4} + \frac{r}{4}\frac{d \xi_r}{dr}.
\end{equation}

Our goal is to solve equations~\eqref{eq:elastic_eom_1}, \eqref{eq:elastic_eom_2}, and \eqref{xi_r_phi_2} to determine the functions $\xi_r$, $\xi_{\perp}$, and $\delta \tilde{\mu}_\mathrm{p}$. We follow a numerical approach similar to that described in \citet{Ushomirsky_2002}, converting the second-order ODEs into first-order form. For this purpose, we introduce the following new variables:
\begin{align}
\label{z_1_xi_r}
    z_1 &= \frac{\xi_r}{r}, \\
    z_2 &= \frac{\Delta \tau_{rr}}{\Psi_{\textcolor{black}{0}}} = \frac{\delta \tau_{rr}}{\Psi_{\textcolor{black}{0}}} - \frac{z_1 }{\Psi_{\textcolor{black}{0}}}\frac{d \Psi}{d \ln r}, \\
    \label{z_3_def}
    z_3 &= \frac{\xi_{\perp}}{r}, \\
    \label{z_4_def}
    z_4 &= \frac{\Delta \tau_{r \phi}'}{\Psi_{\textcolor{black}{0}}} = \frac{\delta \tau_{r \phi}'}{\Psi_{\textcolor{black}{0}}}, \\
    z_5 &= \frac{\Delta \tilde{\mu}_\mathrm{p}}{\tilde{\mu}_o} = \frac{\delta \tilde{\mu}_\mathrm{p}}{\tilde{\mu}_o} +\frac{z_1}{\tilde{\mu}_o}\frac{d  \tilde{\mu}_\mathrm{p}}{d \ln r}.
\end{align}

Here,
\begin{equation}
\label{mu_not}
    \tilde{\mu}_o = C_1 + C_2,
\end{equation}
which is the value of $\tilde{\mu}_\mathrm{p}(r)$ at the origin, and
\begin{equation}
\label{psi_not}
    \Psi_{\textcolor{black}{0}} =  \frac{a_{\mathrm{p}}\tilde{\mu}_o^2}{2} =  \frac{a_{\mathrm{p}}(C_1 + C_2)^2}{2},
\end{equation}
is the central value of $\Psi$. The constant $C_1$ is the same as defined in equation~\eqref{eq:C1_final}, and
\begin{equation}
\label{c2_mu}
    C_2 = \frac{2 \Omega^2}{A}.
\end{equation}

To simplify the analysis, we use the dimensionless radius $\hat{r}$ and three dimensionless parameters:
\begin{equation}
    \varepsilon_{\mu}= \frac{\mu_0}{\Psi_0},
    \label{eq:eps_mu}
\end{equation}
\begin{equation}
    \varepsilon_{\Omega}= \frac{ 2 \pi f^2}{ G (\rho_{n,0} + \rho_{p,0})},
    \label{eq:eps_omega}
\end{equation}
\begin{equation}
    \varepsilon_{\mathrm{\mathrm{Mag}}}= \frac{B_{*} 2 \pi f}{\Psi_0}.
    \label{eq:eps_mag}
\end{equation}

The parameter \( \varepsilon_{\mu} \) measures the relative importance of elastic stresses compared to the background pressure. Smaller values of \( \varepsilon_{\mu} \) correspond to a less stiff crust. The parameter \( \varepsilon_{\Omega} \) quantifies the strength of rotational effects relative to gravity. It can be interpreted as the ratio of centrifugal to gravitational forces, and therefore controls the degree of rotational deformation of the star. The parameter \( \varepsilon_{\mathrm {Mag}} \) characterises the strength of the Magnus force relative to the background pressure. 

Using the definitions of $z_2$ and $z_4$, and substituting from equations~\eqref{tau_rr}, ~\eqref{tau_rperp}, and ~\eqref{psi_pomp}, we obtain two first-order differential equations:
\begin{align}
\label{diff_z1_1}
    \frac{d z_1}{d \ln \hat{r}} &= -\frac{z_1}{2} + \frac{3}{4 \alpha_1}z_2 - 2 z_3 +\frac{3 \alpha_3}{2 \alpha_1}z_5, \\
\label{diff_z3}
    \frac{d z_3}{d \ln \hat{r}} &= \frac{z_4}{\alpha_1} - z_1.
\end{align}

Here,
\begin{equation}
\label{eq:alpha1}
    \alpha_1 = \hat{\rho}^{\textcolor{black}{\frac{4}{3}}} \, \varepsilon_{\mu},
\end{equation}
and
\begin{equation}
\label{eq:alpha3}
    \alpha_3  = \hat{\rho} ,
\end{equation}
where
\begin{equation}
    \hat{\rho} = \frac{\rho_\mathrm{n}}{\rho_{n,0}} = \frac{\rho_\mathrm{p}}{\rho_{p,0}} = \varepsilon_{\Omega} + J_0(\sqrt{A}R \hat{r})(1 - \varepsilon_{\Omega} )
\end{equation}

We now rewrite the EOMs ~\eqref{eq:emo_elastic_r_2_real_pomp}, ~\eqref{eq:eom_elastic_phi_2_real_pomp}, and the continuity equation ~\eqref{xi_r_phi_2} in terms of $z_i$, obtaining:
\begin{align}
\label{diff_z2}
    \frac{d z_2}{d \ln \hat{r}} &= z_1 (\alpha_1 - \frac{\Gamma}{2} - \gamma + \frac{\alpha_4^2}{2}) - z_2\frac{3}{4}\frac{\alpha_2}{\alpha_1} \nonumber \\
&\quad+ 2 z_3 (\Gamma - 2 \alpha_1) + 4 z_4 - \left[\frac{3}{2}\frac{\alpha_2 \alpha_3}{\alpha_1} + \alpha_4 \right] z_5 +\tilde{V}, \\
\label{diff_z4}
    \frac{d z_4}{d \ln \hat{r}} &= - z_1 (\alpha_1 + \Gamma) + \frac{z_2}{2} + 4 \alpha_1 z_3 - 2 z_4 +3 \alpha_3 z_5 + \tilde{U}, \\
\label{diff_z1_2}
    \frac{d z_1}{d \ln \hat{r}} &= -2 z_1 +4z_3 - \frac{z_5}{\alpha_3}.
\end{align}
where
\begin{align}
\label{lambda_pomp}
    \Gamma &= -2 \hat{r} \hat{\rho} (1-\varepsilon_{\Omega}) \sqrt{A} R
    J_1(\sqrt{A} R \hat{r}), \\
\label{alpha_2_pomp}
    \alpha_2 &= \Gamma + 2 \alpha_1, \\
\label{gamma_pomp}
    \gamma &= \hat{r}^2\frac{d^2 (\hat{\rho})^2}{d \hat{r}^2}, \\
    \alpha_4 &= 2\frac{d \hat{\rho}}{dr}, \\
\label{tilde_V}
    \tilde{V} &=  -4 \varepsilon_{\mathrm{\mathrm{Mag}}} \hat{r}^2
    \left(\hat{\rho}^2 + \frac{\Gamma}{2}\right), \\
\label{tilde_U}
    \tilde{U} &= -2 \varepsilon_{\mathrm{\mathrm{Mag}}} \hat{r}^2 \hat{\rho}^2 .
\end{align}

Equating equations~\eqref{diff_z1_1} and ~\eqref{diff_z1_2} allows expressing $z_5$ in terms of the other variables:
\begin{equation}
\label{z_5}
    z_5 = \left( \frac{2  \alpha_3}{3 \alpha_5 +2}\right) \left( - \frac{3}{2} z_1 - \frac{3}{4 \alpha_1} z_2 + 6 z_3\right),
\end{equation} 
where
\begin{equation}
    \alpha_5 = \frac{\alpha_3^2}{\alpha_1} =   \frac{\hat{\rho}^{2/3}}{\varepsilon_{\mu}}.
\end{equation}

Substituting equation~\eqref{z_5} into equations~\eqref{diff_z1_1}, ~\eqref{diff_z3}, ~\eqref{diff_z2}, and ~\eqref{diff_z4} leads to:
\begin{align}
\label{ode_sinle_1}
\frac{d z_1}{d \ln \textcolor{black}{\hat{r}}} &=
- z_1 \left[\frac{6 \alpha_5 + 1}{3 \alpha_5 + 2 }\right]
+ \frac{3 z_2}{2 (3\alpha_3^2 + 2 \alpha_1)} 
+ 4 z_3 \left[\frac{3 \alpha_5 - 1}{3 \alpha_5 + 2}\right],
\\[6pt]
\label{ode_sinle_2}
\frac{d z_2}{d \ln \textcolor{black}{\hat{r}}} &=
z_1 \left[
2 \alpha_1 \left(  \frac{6 \alpha_5 +1}{3 \alpha_5 +2} \right)
+ \Gamma - 2 \alpha_3\beta
\right] 
- z_2 \frac{3}{(3\alpha_5+ 2)} \nonumber \\
&\quad
- z_3 \left[
4 \Gamma
+ 8 \alpha_1
\left(  \frac{6 \alpha_5 +1}{3 \alpha_5 +2} \right)
\right] 
+ 4 z_4 + \tilde{V},
\\[6pt]
\label{ode_sinle_3}
\frac{d z_3}{d \ln \textcolor{black}{\hat{r}}} &=
\frac{z_4}{\alpha_1} - z_1,
\\[6pt]
\label{ode_sinle_4}
\frac{d z_4}{d \ln \textcolor{black}{\hat{r}}} &=
- z_1
\left[
\Gamma
+ \frac{2\alpha_1(6 \alpha_5 + 1)}{3 \alpha_5+ 2}
\right] 
+ z_2
\left[
\frac{1 -3 \alpha_5}{3 \alpha_5+ 2 }
\right]
\nonumber \\
&\quad + 8 \alpha_1 z_3
\left[
\frac{6 \alpha_5 +1 }{3 \alpha_5+ 2 }
\right] 
- 2 z_4 + \tilde{U}.
\end{align}

Here,
\begin{equation}
\label{beta_pomp}
    \beta = - \frac{\hat{r}^2 (\sqrt{A} R)^2 (1-\varepsilon_{\Omega})
    \bigl[J_0(\sqrt{A} R \hat{r}) - J_2(\sqrt{A} R \hat{r})\bigr]}{2}.
\end{equation}

Equations~\eqref{ode_sinle_1}--\eqref{ode_sinle_4} define a coupled system of four first-order ODEs in the variables $z_1$, $z_2$, $z_3$, and $z_4$, where $\tilde{U}$ and $\tilde{V}$ are source terms. 

Having considered perturbations in the crust, we now turn to the fluid core. Since we adopt the Cowling approximation and assume that the velocity perturbation is confined entirely to the crust, there are no perturbations of the form \(\delta f \, e^{im\phi}\) with $m \neq 0$ in the core and ocean regions of the star, where \(f\) denotes a fluid variable. 
%This can be seen explicitly from the perturbed equations of motion (EOMs) in the fluid core and ocean for the neutron superfluid and the charged fluid.  

To see this, recall that the unperturbed equations of motion (EOMs) for the neutron superfluid and the charged fluid are given by equations~\eqref{eq:euler_rotating_1} and~\eqref{eq:euler_rotating_2}, respectively. If we do not adopt the Cowling approximation, then upon perturbation, equations~\eqref{eq:euler_rotating_1} and~\eqref{eq:euler_rotating_2} reduce to
\begin{equation}
\nabla (\delta \mu_X + \delta \Phi) = 0 ,  \quad X \in \{n, p\},
\label{eq:perturbed_eom_rot_3}
\end{equation}
where \(\delta \mu_X\) is the perturbed chemical potential for each component and \(\delta \Phi\) is the perturbed gravitational potential. Since we consider the perturbed velocity field only in the crust, there are no velocity perturbations in the core or ocean.  
In this case, the perturbed gravitational potential exactly balances the perturbed chemical potential for each component, and therefore \(\delta \mu_X\) can admit perturbation components with $m \neq 0$.  

If we instead adopt the Cowling approximation, then upon perturbation, equations~\eqref{eq:euler_rotating_1} and~\eqref{eq:euler_rotating_2} reduce to
\begin{equation}
\nabla (\delta \mu_X) = 0 , 
\label{eq:perturbed_eom_rot_4}
\end{equation}
which follows directly from the absence of velocity perturbations together with the neglect of the perturbed gravitational potential. We can see that the Cowling approximation introduces a significant difference: under this assumption, \(\delta \mu_X\) cannot contain perturbation components with $m \neq 0$.  
An essentially identical argument, formulated in terms of the pressure perturbation, was presented by \citet{Ushomirsky_2002}.  
Therefore, for the \(m=2\) source term, we solve the perturbed equations of motion only within the annular crustal region.  

We have allowed for pinning only in the solid crust. One could, however, also consider pinning both to crustal nuclei and to magnetic flux tubes in the core, as illustrated in Fig.~\ref{pinn}. In this scenario, the neutron superfluid velocity perturbation exists only in the crust and the core and vanishes in the ocean. The modified ansatz for the streamfunction \(\textcolor{black}{\mathcal{S}}\), chosen to ensure that the perturbation vanishes at the crust--ocean boundary, is given by:

\begin{equation}
\label{psi_annulus_2}
    \textcolor{black}{\mathcal{S}} = \frac{r^2}{R^2} \left(\frac{r^2}{R_{\text{out}}^2} - 1\right) e^{i2\phi}.
\end{equation}

Substituting equation~\eqref{psi_annulus_2} into the general expression for the velocity perturbation, as given in equation~\eqref{eq:pert_v_def_final}, we obtain:

\begin{equation}
    \vec{\delta v_\mathrm{n}} = -\frac{2i B_*}{\rho_\mathrm{n}}\frac{r}{R^2}\left( \frac{r^2}{R_{\text{out}}^2} - 1\right) e^{2i\phi} \hat{r} 
    + \frac{B_*}{\rho_\mathrm{n}}\left( \frac{4 r^3}{R^2 R_{\text{out}}^2} - 2 \frac{r}{R^2} \right)e^{2i\phi} \hat{\phi}.
\end{equation}

This expression gives the perturbation to the neutron superfluid velocity field. One can then attempt to solve the perturbed equations of motion in the core. In the core, the proton fluid experiences a reaction force that is equal in magnitude and opposite in direction to the Magnus force. The perturbed equation of motion for the proton fluid is

\begin{equation}
\label{proton_eom}
    \vec{\nabla} \delta \tilde{\mu}_{\mathrm{p}}= \frac{\rho_\mathrm{n}}{\rho_\mathrm{p}} \, 2 \vec{\Omega} \times \vec{\delta v_\mathrm{n}}.
\end{equation}

Equation~\eqref{proton_eom} is obtained by perturbing the unperturbed fluid EOM, originally given in equation~\eqref{eq:euler_rotating_2}. In this equation, the left-hand side is the gradient of a scalar, whereas the right-hand side involves the Coriolis force term, which is not expressible as a gradient of a scalar field.  In general, these forces cannot balance each other. Dropping the Cowling approximation would not help in this case, as the inclusion of gravitational potential terms would still result in scalar gradients and hence not contribute to the non-conservative force required on the right-hand side.

\textcolor{black}{In the present calculation, we assume that the perturbed fluids obey the same barotropic equations of state as the unperturbed background. More generally, however, if the perturbation timescale is short compared with the timescale on which weak interactions restore beta equilibrium, the perturbative equation of state can differ from the equilibrium one, giving rise to a non-zero Schwarzschild-discriminant term in the EOM (see, e.g., \citep{AnderssonComer2001}). Such an additional buoyancy term would modify the perturbed equation of motion and hence the corresponding integrability condition. A consistent treatment of this effect would require including the perturbation and beta-equilibration timescales explicitly, which lies beyond the scope of the present model.}

\textcolor{black}{In addition to this possible non-barotropic contribution, other physical effects may also modify the force balance.}  We suggest that in a more realistic neutron star, the reaction force on the proton fluid might also be counteracted by magnetic tension forces associated with flux tubes. However, such magnetic effects are beyond the scope of the present simplified model. We need a more detailed investigation in future studies, possibly involving the inclusion of magnetic field effects in the core.

\section{Boundary conditions}
\label{bc}

To solve the four first-order ODEs of  equations~\eqref{ode_sinle_1}--\eqref{ode_sinle_4}, we require four boundary conditions. Physical considerations lead to two boundary conditions at the crust-core interface and two at the crust-ocean interface, as outlined below.
\begin{enumerate}
    \item At the crust-core boundary ($r = R_{\text{in}}$):
    \begin{enumerate}
        \item The first condition enforces the vanishing of the Lagrangian perturbation of the tangential traction component in both the fluid and elastic regions at the interface:
        \begin{equation}
        \label{bc_1_annulus}
            \Delta \tau_{r \phi}^E  =  \Delta \tau_{r \phi}^F = 0,
        \end{equation}
        where the superscripts \(E\) and \(F\) refer to the elastic and fluid components, respectively. This can be written in terms of the $z_4$ variable as:
        \begin{equation}
        \label{bc_ann_1}
            z_4 = 0.
        \end{equation}

        \item The second condition involves the Lagrangian perturbation of the radial traction component at the interface:
        \begin{equation}
        \label{bc_2_annulus22}
           \Delta \tau_{r r}^E =  \Delta \tau_{r r}^F.
        \end{equation}
        For the fluid component, the Lagrangian perturbation of the radial traction can be expressed as
        \begin{equation}
        \label{bc_2_annulus22_2}
           \Delta \tau_{r r}^F =   \delta \tau_{r r}^F + \xi_F^r \nabla_r \tau_{r r}^F,
        \end{equation}
        where the background stress tensor \(\tau_{r r}^F\) is given by 
        \begin{equation}
        \label{tau_rr_f}
            \tau_{r r}^F =  - \frac{a_{\mathrm{p}}\tilde{\mu_\mathrm{p}}^2}{2},
        \end{equation}
        following the same reasoning used in the derivation of equation~\eqref{stress_back_pomp}.
        
        Taking the Eulerian perturbation of equation~\eqref{tau_rr_f} yields
        \begin{equation}
        \label{eulerian_stress}
            \delta \tau_{r r}^F  = - a_{\mathrm{p}}\delta \tilde{\mu_\mathrm{p}}^F.
        \end{equation}

        Substituting equation~\eqref{eulerian_stress} into equation~\eqref{bc_2_annulus22_2} and applying the condition from equation~\eqref{bc_2_annulus22}, we obtain
        \begin{equation}
           \Delta \tau_{r r}^E =  - a_{\mathrm{p}}\delta \tilde{\mu_\mathrm{p}}^F + \xi^r_F \nabla_r\tau_{r r}^F.
        \end{equation}

        Since \(\xi^r_F = \xi^r_E\) (the radial displacement must be the same at the interface for both components) and \(\delta \tilde{\mu_\mathrm{p}}^F = 0\) at the crust–core interface (due to the absence of \(m \neq 0\) perturbations in the fluid core), this simplifies to

        \begin{equation}
        \label{bc_2_annulus_2}
           \Delta \tau_{r r}^E =  \xi^r_E \nabla_r\tau_{r r}^F.
        \end{equation}

        Inserting (\ref{tau_rr_f}) into equation~\eqref{bc_2_annulus_2} and differentiating gives:
        \begin{equation}
        \label{bc_2_annulus}
           \Delta \tau_{r r}^E =  - \xi^r_E a_{\mathrm{p}}\tilde{\mu_\mathrm{p}} \frac{d \tilde{\mu_\mathrm{p}}}{dr}\bigg|_{r = R_{\text{in}}}.
        \end{equation}

        Dividing both sides by $\Psi_{\textcolor{black}{0}}$ yields:
        \begin{equation}
        \label{bc_2_annulus_3}
           \frac{\Delta \tau_{r r}^E}{\Psi_{\textcolor{black}{0}}} =  - \frac{\xi^r_E a_{\mathrm{p}}\tilde{\mu_\mathrm{p}}}{\Psi_{\textcolor{black}{0}}} \frac{d \tilde{\mu_\mathrm{p}}}{dr}\bigg|_{r = R_{\text{in}}}.
        \end{equation}

        This can be written in terms of the variables $z_1$ and $z_2$ as:
        \begin{equation}
        \label{bc_2_annulus_final}
           z_2 =  - z_1 \frac{ R_{\text{in}} a_{\mathrm{p}}\tilde{\mu_\mathrm{p}}}{\Psi_{\textcolor{black}{0}}} \frac{d \tilde{\mu_\mathrm{p}}}{dr}\bigg|_{r = R_{\text{in}}}.
        \end{equation}
    \end{enumerate}

    These boundary conditions, given by Equations~\eqref{bc_ann_1} and~\eqref{bc_2_annulus_final}, are similar to those presented in equation~(47) of \citet{Ushomirsky_2002}. In their case, however, the analysis was carried out for spherical stars.

    \item Similarly, at the crust-ocean boundary ($r = R_{\text{out}}$):
    \begin{enumerate}
        \item The tangential traction must vanish:
        \begin{equation}
        \label{bc_3_annulus}
            z_4 = 0.
        \end{equation}

        \item The radial traction condition is:
        \begin{equation}
        \label{bc_4_annulus_final}
           z_2 =  - z_1 \frac{ R_{\text{out}} a_{\mathrm{p}}\tilde{\mu_\mathrm{p}}}{\Psi_{\textcolor{black}{0}}} \frac{d \tilde{\mu_\mathrm{p}}}{dr}\bigg|_{r = R_{\text{out}}}.
        \end{equation}
    \end{enumerate}
\end{enumerate}

So we do indeed have  four boundary conditions, two at the crust-core interface and two at the crust-ocean interface, sufficient to solve the system of four coupled first-order ODEs.

\section{Quadrupole ellipticity}
\label{quad}
To obtain a global measure of the stellar deformation, we calculate the quadrupole ellipticity, defined as

\begin{equation}
\label{eq:quad_def}
\epsilon_Q
\equiv
\frac{I_{xx}-I_{yy}}{I_{zz}}.
\end{equation}
This quantity is the cylindrical analogue of the ellipticity commonly used for spherical stars and is the quantity that enters the quadrupole formula for gravitational-wave emission. \textcolor{black}{Although its definition is analogous to the equatorial ellipticity used for spherical neutron stars, its numerical value cannot be directly identified with that of a spherical star. The cylindrical geometry differs from the spherical geometry in its moment-of-inertia distribution. A recent spherical calculation by \citet{Giliberti2026} finds Magnus-mountain ellipticities more than an order of magnitude smaller than those obtained in our cylindrical model. We discuss this comparison further in Section~\ref{summary_ponclusion}. Our results should therefore be interpreted as estimates within the present cylindrical model, with the main aim of understanding how the Magnus force gives rise to stellar deformation.}

For our cylindrical model, the moment of inertia about the rotation axis is given by
\begin{equation}
\label{eq:Izz}
I_{zz}
=
2\pi L \int_0^R (\rho_{\mathrm{p}}+ \rho_\mathrm{n})\, r^3\, dr,
\end{equation}
where \( L \) is the length of the cylinder.

Since the background configuration is axisymmetric, only the perturbations contribute to the difference \( I_{xx} - I_{yy} \), so that
\begin{equation}
\label{eq:deltaI_difference}
I_{xx} - I_{yy}
=
\delta I_{xx} - \delta I_{yy}.
\end{equation}
Substituting equation~\eqref{eq:deltaI_difference} into equation~\eqref{eq:quad_def}, we obtain
\begin{equation}
\label{eq:quad_perturbed}
\epsilon_Q
=
\frac{\delta I_{xx} - \delta I_{yy}}{I_{zz}}.
\end{equation}

Following \citet{yim}, the perturbation to the inertia tensor is given by
\begin{equation}
\label{eq:deltaI_general}
\delta I_{ij}
=
\int_V \rho(\mathbf{x})
\left(
\xi_i x_j + x_i \xi_j
\right)\, dV.
\end{equation}

For the \( xx \)-component, equation~\eqref{eq:deltaI_general} gives
\begin{equation}
\label{eq:deltaI_xx_step1}
\delta I_{xx}
=
\int_V \rho(\mathbf{x})
\left(
\xi_x x + x \xi_x
\right)\, dV
=
2 \int_V \rho(\mathbf{x})\, x\, \xi_x \, dV.
\end{equation}
Transforming to cylindrical coordinates yields
\begin{equation}
\label{eq:deltaI_xx_pyl}
\delta I_{xx}
=
2 \int
\rho(r)
\left[
\xi_r \cos^2\phi
-
\xi_\phi \sin\phi \cos\phi
\right]
r^2\, dr\, d\phi\, dz.
\end{equation}
Using equations~\eqref{z_1_xi_r} and \eqref{z_3_def}, this becomes
\begin{multline}
\label{eq:deltaI_xx_sub}
\delta I_{xx} = 2 \int \rho(r) \Bigl[
    z_1 \cos(2\phi)\cos^2\phi \\
    {}+ 2z_3 \sin(2\phi)\sin\phi\cos\phi
\Bigr] r^3\, dr\, d\phi\, dz.
\end{multline}
Integrating over the cylinder length \( L \) in the \( z \)-direction and over the azimuthal angle \( \phi \), we obtain
\begin{equation}
\label{eq:deltaI_xx_final}
\delta I_{xx}
=
\pi L \int \rho(r)\left[z_1 + 2z_3\right] r^3\, dr.
\end{equation}
Similarly, for the \( yy \)-component, we obtain
\begin{equation}
\label{eq:deltaI_yy_final}
\delta I_{yy}
=
-\pi L \int \rho(r)\left[z_1 + 2z_3\right] r^3\, dr.
\end{equation}

\textcolor{black}{Although the neutron superfluid possesses a non-axisymmetric velocity perturbation, it does not contribute to the bulk mass quadrupole at linear order in our model. As shown in Section~\ref{Solving Elastic Equations of Motion}, the cancellation of the Coriolis and Magnus forces gives $\delta\tilde\mu_{\mathrm n}=0$. Using the polytropic equation of state, this implies $\delta\rho_{\mathrm n}=0$. Furthermore, the neutron radial displacement vanishes at the stellar surface under our adopted boundary condition. Consequently, the mass quadrupole is determined entirely by the charged-component deformation.}

Substituting equations~\eqref{eq:deltaI_xx_final} and \eqref{eq:deltaI_yy_final} into equation~\eqref{eq:quad_perturbed}, we obtain
\begin{equation}
\label{eq:quad_exp_1}
\epsilon_Q
=
\frac{
\displaystyle \int_{R_{\rm in}}^{R_{\rm out}}
\rho_\mathrm{p}(r)\left[z_1(r) + 2z_3(r)\right] r^3\,dr
}{
\displaystyle \int_0^R (\rho_\mathrm{p}(r) + \rho_\mathrm{n}(r))\, r^3\,dr
}.
\end{equation}
From equation~\eqref{lambda_ratio}, we have
\begin{equation}
\label{eq:density_ratio}
\frac{\rho_\mathrm{p}}{\rho_\mathrm{n}} = \frac{1}{9}.
\end{equation}
Using equation~\eqref{eq:density_ratio}, equation~\eqref{eq:quad_exp_1} simplifies to
\begin{equation}
\label{eq:quad_exp_2}
\epsilon_Q
=
\frac{
\displaystyle \int_{R_{\rm in}}^{R_{\rm out}}
\rho_\mathrm{p}(r)\left[z_1(r) + 2z_3(r)\right] r^3\,dr
}{
\displaystyle 10 \int_0^R \rho_\mathrm{p}(r)\, r^3\,dr
}.
\end{equation}

Dividing both the numerator and denominator by the central
charged-fluid density \( \rho_{p,0} \),
\textcolor{black}{expressing the integrals in terms of the
dimensionless radial coordinate $\hat{r}$,}
and using equation~\eqref{eq:alpha3}, we obtain

\begin{equation}
\label{eq:quad_final}
\epsilon_Q =
\frac{
\displaystyle \int_{0.9}^{0.99}
\alpha_3(\hat{r})
\left[z_1(\hat{r})+2z_3(\hat{r})\right]
\hat{r}^3\,d\hat{r}
}{
\displaystyle 10\int_0^1
\alpha_3(\hat{r})\,\hat{r}^3\,d\hat{r}
}
\end{equation}

Thus, for a given set of parameters \( \varepsilon_{\mu} \), \( \varepsilon_{\Omega} \), and \( \varepsilon_{\mathrm{\mathrm{Mag}}} \), equation~\eqref{eq:quad_final} gives the effective mountain size. In a later section, we will use equation~\eqref{eq:quad_final} to calculate the corresponding quadrupole ellipticity.

\section{Parametrised dimensionless quantities and back-of-the-envelope estimates}
\label{BOTE}

In this section, we write the three dimensionless parameters $\varepsilon_{\mu}$, $\varepsilon_{\Omega}$, and $\varepsilon_{\rm Mag}$, as defined in equations~\eqref{eq:eps_mu}, \eqref{eq:eps_omega}, and \eqref{eq:eps_mag}, in parametrised form.  We also make a back-of-the-envelope (BOTE) estimate of the crustal displacement variables $z_1$ and $z_3$, writing this estimate in terms of the dimensionless parameters.

Using the relation $\rho_{p,0}=\rho_{n,0}/9$, so that $\rho_{n,0}+\rho_{p,0}=(10/9)\rho_{n,0}$, the rotational parameter may be written in the parametrised form
\begin{equation}
\varepsilon_{\Omega} 
\simeq  10^{-3}\,f_{100}^2\,\rho_{15}^{-1} ,
\end{equation}
where $f_{100}=f/(100~\mathrm{Hz})$ and
$\rho_{15}=\rho_{n,0}/(10^{15}~\mathrm{g\,cm^{-3}})$.

We identify the amplitude of the velocity perturbation, $\delta v_\mathrm{n}$, \textcolor{black}{measured in the rotating frame,} with the critical relative velocity between the neutron superfluid and the charged component immediately prior to vortex unpinning.  \citet{Pizzo2011} computed the critical angular-velocity lag profile using the snowplow model 
%for an $n=1$ polytropic $1.4\,M_\odot$ neutron star, 
obtaining a maximum critical lag of approximately $\Delta\Omega_{\rm crit}\sim10^{-2}~\mathrm{rad\,s^{-1}}$. A similar critical-lag profile was obtained by \citet{critical_lag} using realistic mesoscopic pinning-force calculations. 
%for a $1.4\,M_\odot$ neutron-star model constructed by solving the Tolman--Oppenheimer--Volkoff (TOV) equations with the GM1 equation of state (their Fig.~15). 
Using
\begin{equation}
\delta v_{\mathrm{n}}\approx r\,\Delta\Omega_{\rm crit},
\end{equation}
and evaluating the expression at \(r = 0.95 \times 10^6\,\mathrm{cm}\), corresponding to the midpoint of the crust in our model, gives
\[
\delta v_{\mathrm{n}}\approx (0.95\times10^6~\mathrm{cm})(10^{-2}~\mathrm{s^{-1}})
\approx 10^4~\mathrm{cm\,s^{-1}},
\]
which we adopt throughout this work.  All of our perturbation results scale (exactly) linearly in $\delta v_{\mathrm{n}}$, so the reader can easily rescale to other critical unpinning velocities.

Using equation~\eqref{eq:pert_v_def_final}, which relates $\delta v_\mathrm{n}$ and $B_*$, we can write the approximate magnitude of the neutron velocity perturbation as
\begin{equation}
\label{vn_mid_1}
\delta v_{\mathrm{n}}\approx \frac{2  \times 10^{-2} B_* }{\rho_{n}(r=0.95R)\, R} \approx \frac{2  \times 10^{-1} B_* }{\rho_{n,0}\, R},
\end{equation}
where we have taken
$\rho_\mathrm{n}(r=0.95R) \approx 0.1 \times \rho_{n,0}$.
This leads to
\begin{equation}
\label{ep_mag_para}
\varepsilon_{\rm Mag} 
=\frac{2 \pi \rho_{n,0}Rf\,\delta v_\mathrm{n}}{2 \times10^{-1}\, \Psi_0}
\simeq 3.14 \times 10^{-7}\,
\frac{\rho_{15}\,R_6\,f_{100}\,(\delta v_\mathrm{n})_4}{\Psi_{35}},
\end{equation}
where $R_6=R/(10^6~\mathrm{cm})$, $(\delta v_\mathrm{n})_4=\delta v_\mathrm{n}/(10^4~\mathrm{cm\,s^{-1}})$, and
$\Psi_{35}=\Psi_0/(10^{35}~\mathrm{erg\,cm^{-3}})$. For an elastically relaxed background star, \(\Psi\) is analogous to the \textcolor{black}{central pressure of the charged component, which is \(4 \times 10^{34}\,\mathrm{erg\,cm^{-3}}\)  for our fiducial model, as given in Appendix~\ref{Back_env}, equation~\eqref{eq:central_pressure_p}. This is approximately one order of magnitude smaller than the total central pressure given in equation~\eqref{eq:central_pressure_tot}. We use \(10^{35}\,\mathrm{erg\,cm^{-3}}\) as a convenient order-of-magnitude normalisation scale for the parametrisation, corresponding to \(\Psi_{35}\simeq0.4\) for our fiducial model.}

Finally, the elastic parameter may be written as
\begin{equation}
\label{ep_mu_para}
\varepsilon_{\mu}  =10^{-5}\,\frac{\mu_{30}}{\Psi_{35}},
\end{equation}
with $\mu_{30}= \mu_0/(10^{30}~\mathrm{erg\,cm^{-3}})$. The reference value \(\mu_0=10^{30}~\mathrm{erg\,cm^{-3}}\) is representative of the shear modulus expected in the inner crust of a neutron star \citep{Ushomirsky_2002}.  More speculatively, neutron stars might contain solid cores, possibly composed of quarks, or exotic high-density QCD phases.  In such a scenario, the elasticity parameter could be as large as $\varepsilon_{\mu} \sim 0.1$; see e.g.\ \citet{owen_05, hetal_07}.  
  
Next, we present a back-of-the-envelope estimate of the dimensionless displacement fields \( z_1 \) and \( z_3 \), by balancing the Magnus force with the elastic restoring force. The Magnus force per unit volume is given in equation~\eqref{eq:magnus_linear}.  The elastic restoring force per unit volume is approximated by $\mu \nabla^2 \xi$ (see e.g. Eqs. (\ref{eq:eom_elastic_pomp_3}) and (\ref{eq:pert_stress_tensor_pom})).  In our model, there are spatial derivatives along the radial direction with characteristic lengthscale equal to the crustal thickness $\Delta R$, and along the tangential directions with lengthscale of order $R$.  We therefore approximate this force per unit volume as $\mu \xi / (R \Delta R)$.  By equating the Magnus and elastic forces we obtain
\begin{equation}
\label{app_force_balance}
2 \rho_{n,0}\, \vec{\Omega} \times \vec{v}_\mathrm{n}
=  \mu_0 \frac{\vec{\xi}}{R \Delta R},
\end{equation}
Solving equation~\eqref{app_force_balance} for the magnitude of the displacement field \( \vec{\xi} \), we find 
\begin{equation}
\label{z1_est}
z_1 \approx z_3 
\approx \frac{\xi}{R}
\approx \frac{2 \times 10^{-1}\, \Omega \,  \rho_{n,0}\, v_{\mathrm{n}}\, R}{\mu_0} 
%  \left(      \frac{\Delta R/R}{0.1}   \right),
\end{equation}
where we have taken the crustal thickness to be $\Delta R=1\,\mathrm{km}$, corresponding to
$\Delta R/R\simeq0.1$ for a star of radius $R=10\,\mathrm{km}$.
\begin{figure}
    \centering
    \includegraphics[width=1\columnwidth]{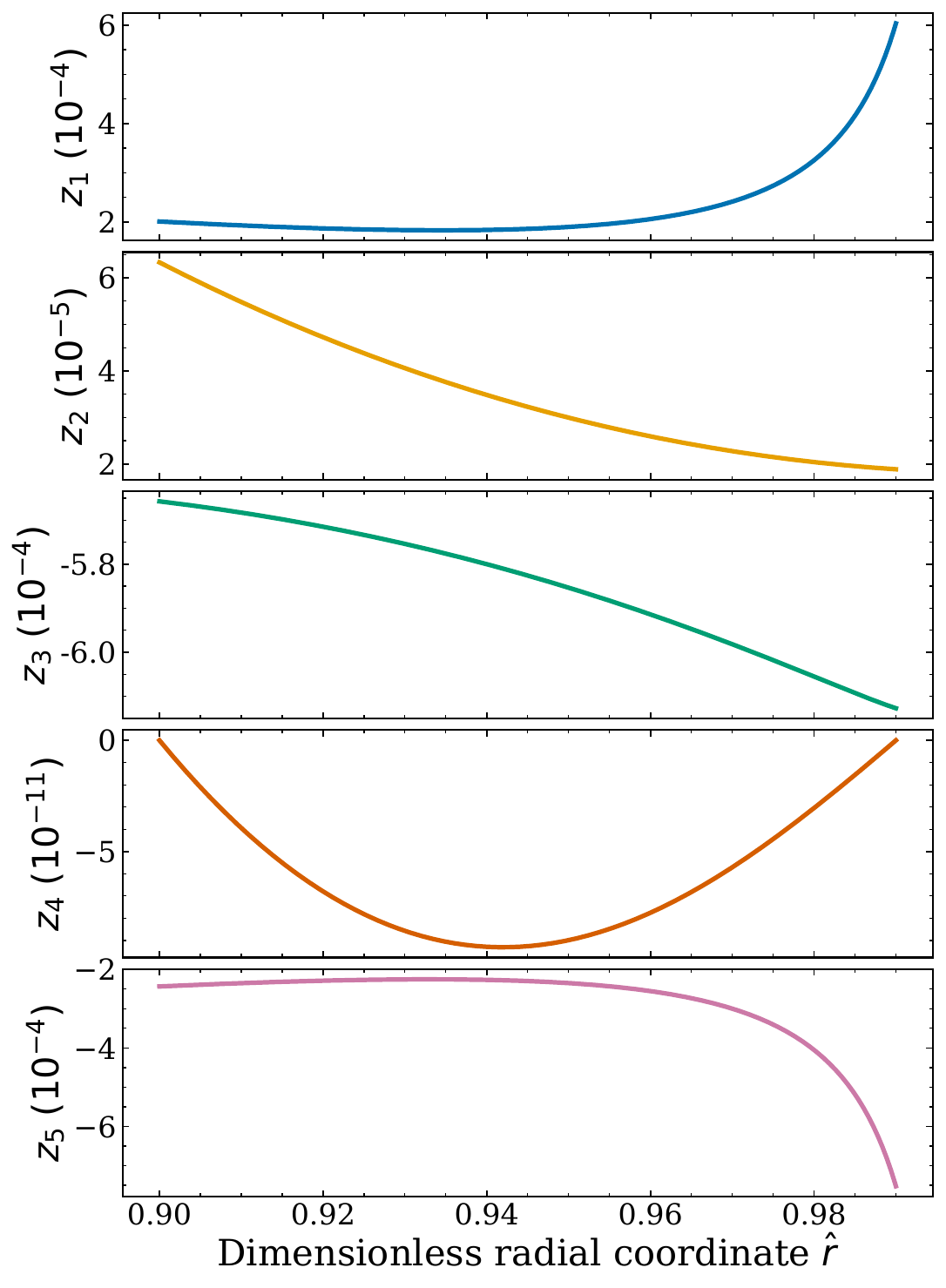}
    \caption{Solutions for $z_1(\hat{r})$, $z_2(\hat{r})$, $z_3(\hat{r})$, $z_4(\hat{r})$ and $z_5(\hat{r})$ as functions of the cylindrical dimensionless radial coordinate $\hat{r}$. The results are shown for the fiducial parameter values \(\varepsilon_{\mu}=10^{-5}\), \(\varepsilon_{\Omega}=10^{-3}\), corresponding to $f_{\mathrm{spin}}=100\,\mathrm{Hz}$, and \(\varepsilon_{\mathrm{Mag}}=10^{-7}\)}.
    \label{fig:R5}
\end{figure}
Substituting equation~\eqref{vn_mid_1} and suppressing the dependence on $\Delta R / R$ we obtain
\begin{equation}
\label{z1_est_1}
z_1 \approx z_3 
 \approx \frac{4 \times 10^{-2}\Omega \,  B_* }{   \, \mu_0} ,
 %\left(   \frac{\rho_{\mathrm{n}}/\rho_{n,0}}{0.1}  \right)
    %\left(   \frac{\Delta R/R}{0.1}   % \right),
\end{equation}
where we have taken $\rho_\mathrm{n}(r=0.95R)/\rho_{n,0} \approx 0.1$. 

Multiplying both the numerator and denominator of equation~\eqref{z1_est_1} by \( \Psi_0 \), and using the definitions of the dimensionless parameters \( \varepsilon_{\mathrm{Mag}} \) and \( \varepsilon_{\mu} \), we can write our estimate in terms of the dimensionless quantities introduced above:
\begin{equation}
\label{z1_est_3}
z_1 \approx z_3 
\approx 4  \times 10^{-2} \frac{ \varepsilon_{\mathrm{Mag}}}{ \varepsilon_{\mu}} .
 %\left( \frac{\rho_{\mathrm{n}}/\rho_{n,0}}{0.1} \right) 
 %\left( \frac{\Delta R/R}{0.1} \right) .
\end{equation}
Converting back to dimensional parameters:
%Finally, for \( \varepsilon_{\mathrm{Mag}} =  3.14 \times 10^{-7} \) and \( \varepsilon_{\mu} = 10^{-5} \), 
equation~\eqref{z1_est_3} gives
\begin{equation}
\label{z1_est_final}
    z_1 \approx z_3 \approx 10^{-3} 
   % \left( \frac{\rho_{\mathrm{n}}/\rho_{n,0}}{0.1} \right) 
    %\left( \frac{\Delta R/R}{0.1} \right) 
    \frac{ \rho_{15}R_6f_{100} (\delta v_{\mathrm{n}})_4 }{ \mu_{30}} .
\end{equation}
The linear scaling in spin frequency $f$ is a direct consequence of the frequency scaling (at fixed $\delta v_{\mathrm{n}}$) of the Magnus force. The scaling  with $\varepsilon_{\mu}^{-1}$ (or equivalently with $\mu^{-1}$) also has a simple interpretation: a stiffer crust provides a stronger elastic restoring force and is therefore more resistant to deformation.

There is no obvious way to make corresponding \textcolor{black}{back-of-the-envelope} (BOTE) estimates for the remaining dependent variables ($z_2, z_4, z_5$), so we do not attempt to do so.

The estimates of Eqs.(\ref{z1_est_3}) and (\ref{z1_est_final}) will serve as useful cross-checks for validating the numerical solutions obtained for the displacement variables \( z_1 \) and \( z_3 \) in the following section.

\section{Numerical results}
\label{NR}

We now numerically solve the system of four first-order ODEs for \(z_i\) (\(i=1,2,3,4\)) described in equations~\eqref{ode_sinle_1}--\eqref{ode_sinle_4}, using the boundary conditions given in equations~\eqref{bc_1_annulus}--\eqref{bc_4_annulus_final}.  The variable \(z_5\) is then obtained algebraically from equation~\eqref{z_5}. The domain of integration is that of the crust, from $\hat{R}_{\text{in}} = 0.9$ to $\hat{R}_{\text{out}} =0.99$, as explained in Section \ref{Solving Elastic Equations of Motion}.

To illustrate the behaviour of the solutions, in Fig.~\ref{fig:R5} we show the numerical solutions for $z_1(\hat{r})$, $z_2(\hat{r})$, $z_3(\hat{r})$, $z_4(\hat{r})$ and $z_5(\hat{r})$ as functions of the cylindrical dimensionless radial coordinate $\hat{r}$.  For the plots, we have adopted the fiducial parameter values \(\varepsilon_{\mu}=10^{-5}\), \(\varepsilon_{\Omega}=10^{-3}\), and \(\varepsilon_{\mathrm{\mathrm{Mag}}}=10^{-7}\), motivated by the order-of-magnitude estimates presented in Section~\ref{BOTE}.   All five solutions are smooth and well behaved across the entire radial domain.

\begin{figure*}
\centering

% Row 1
\begin{minipage}{0.495\textwidth}
\centering
\includegraphics[width=\linewidth]{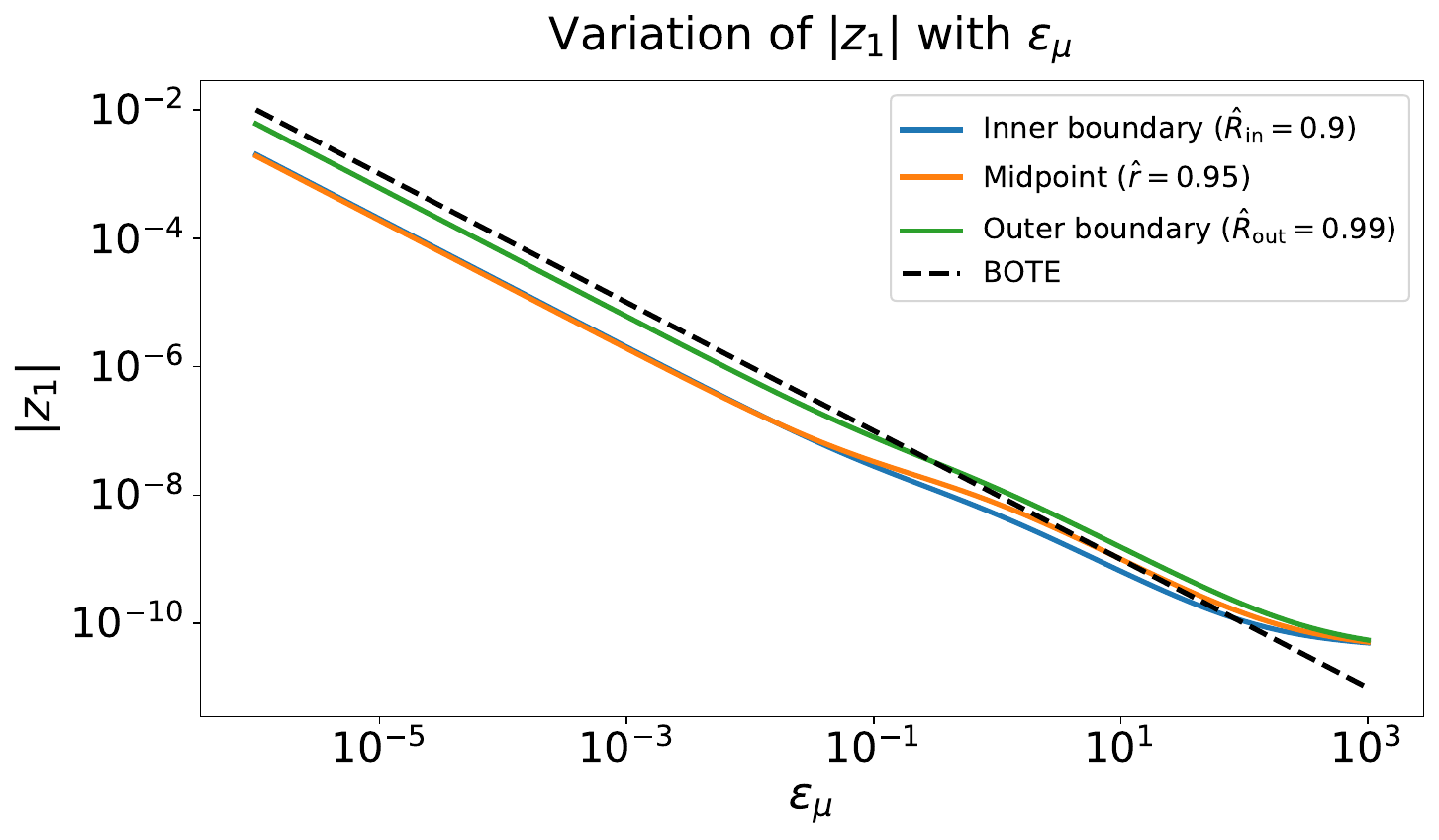}
\\[2mm]
(a)
\end{minipage}\hfill
\begin{minipage}{0.495\textwidth}
\centering
\includegraphics[width=\linewidth]{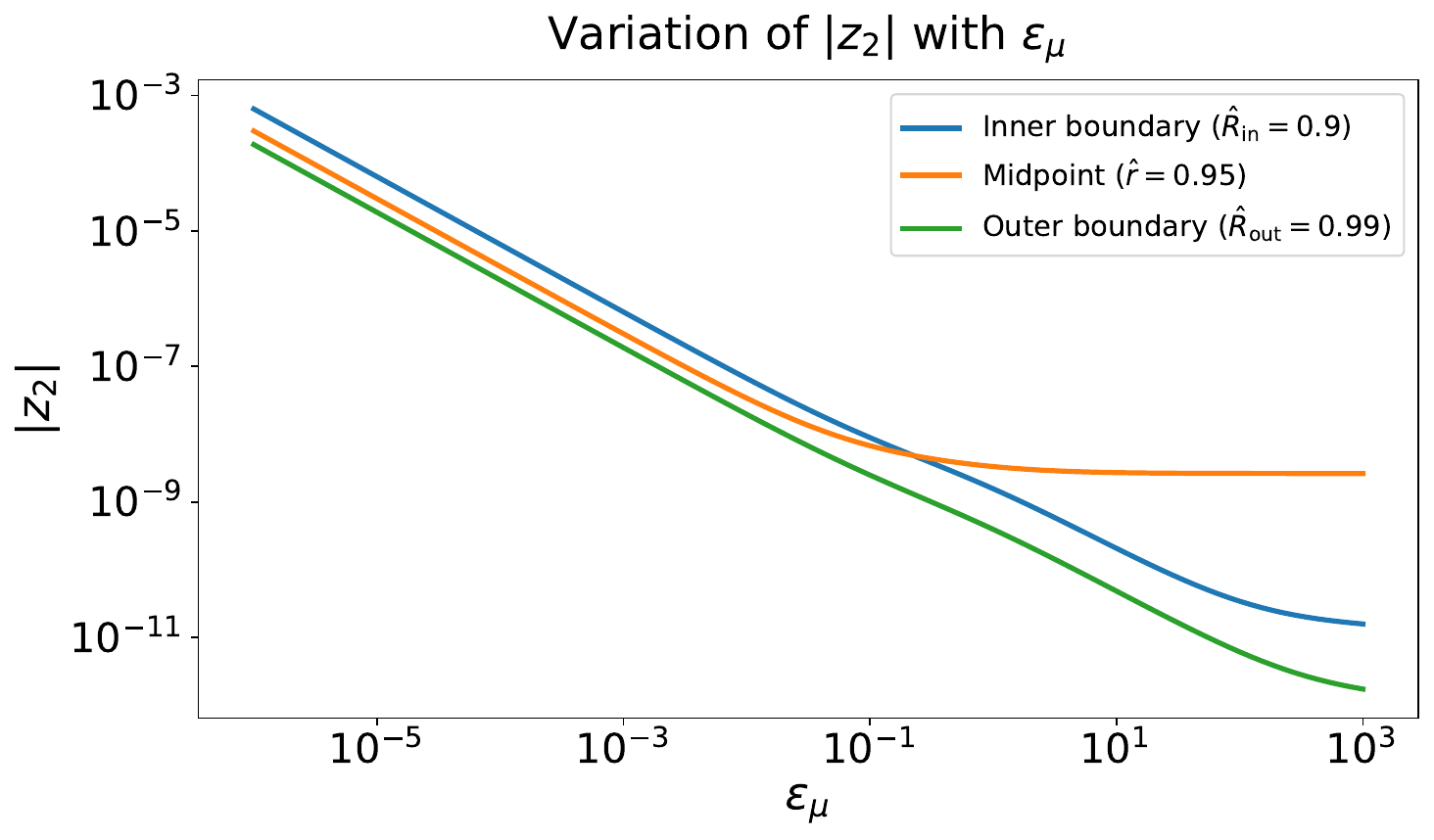}
\\[2mm]
(b)
\end{minipage}

\vspace{0.3cm}

% Row 2
\begin{minipage}{0.495\textwidth}
\centering
\includegraphics[width=\linewidth]{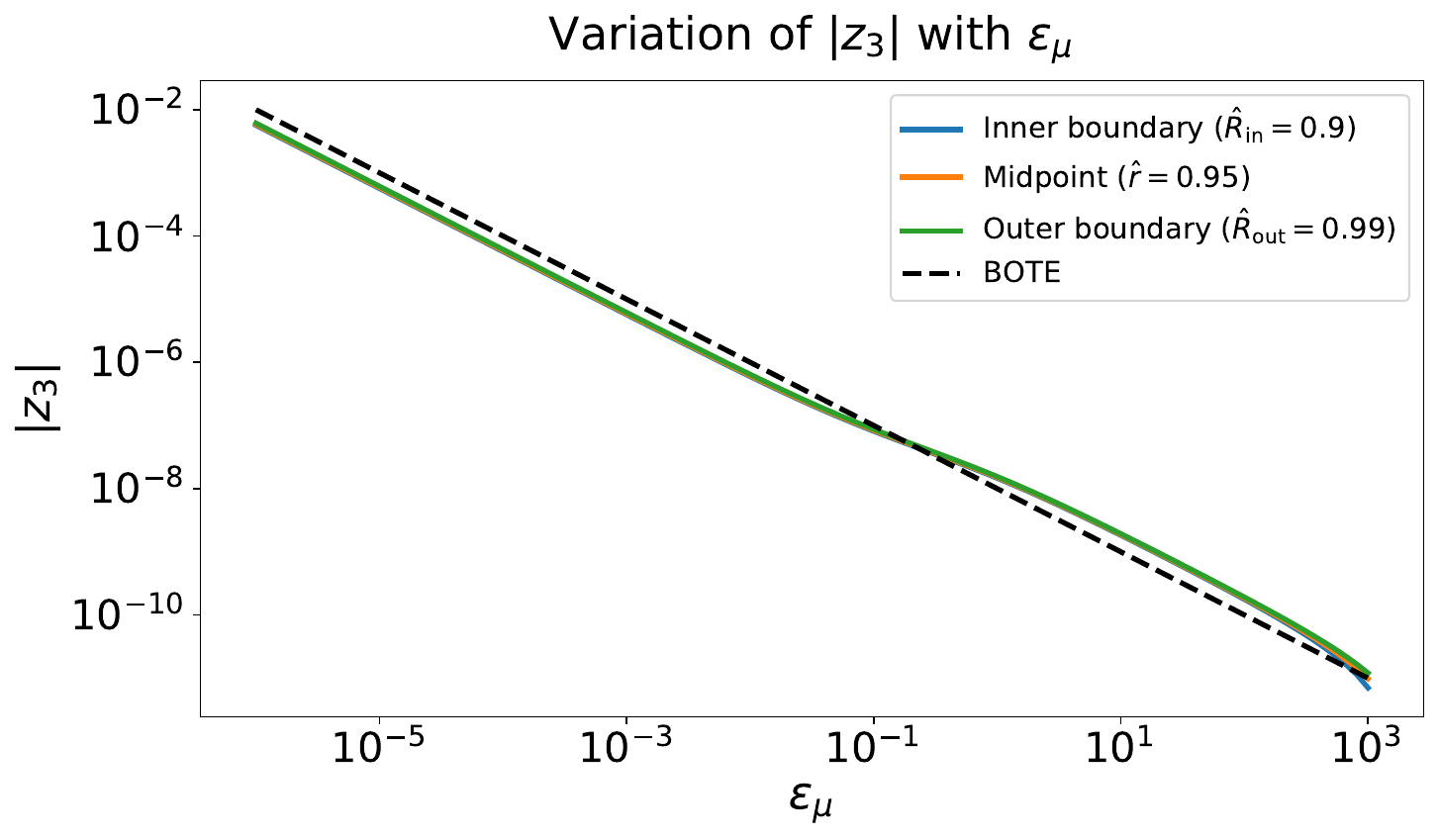}
\\[2mm]
(c)
\end{minipage}\hfill
\begin{minipage}{0.495\textwidth}
\centering
\includegraphics[width=\linewidth]{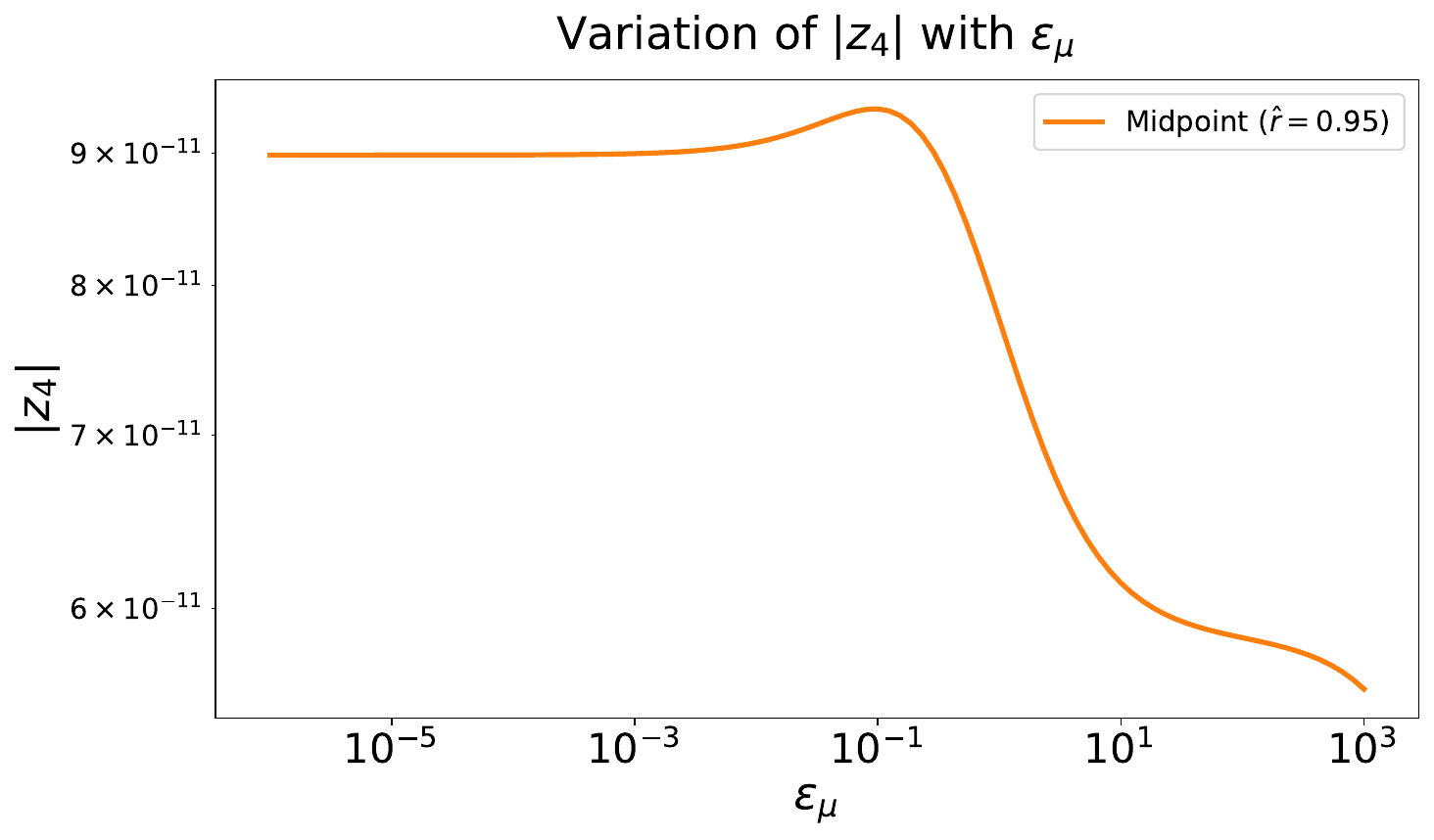}
\\[2mm]
(d)
\end{minipage}

\vspace{0.3cm}

% Row 3
\begin{minipage}{0.495\textwidth}
\centering
\includegraphics[width=\linewidth]{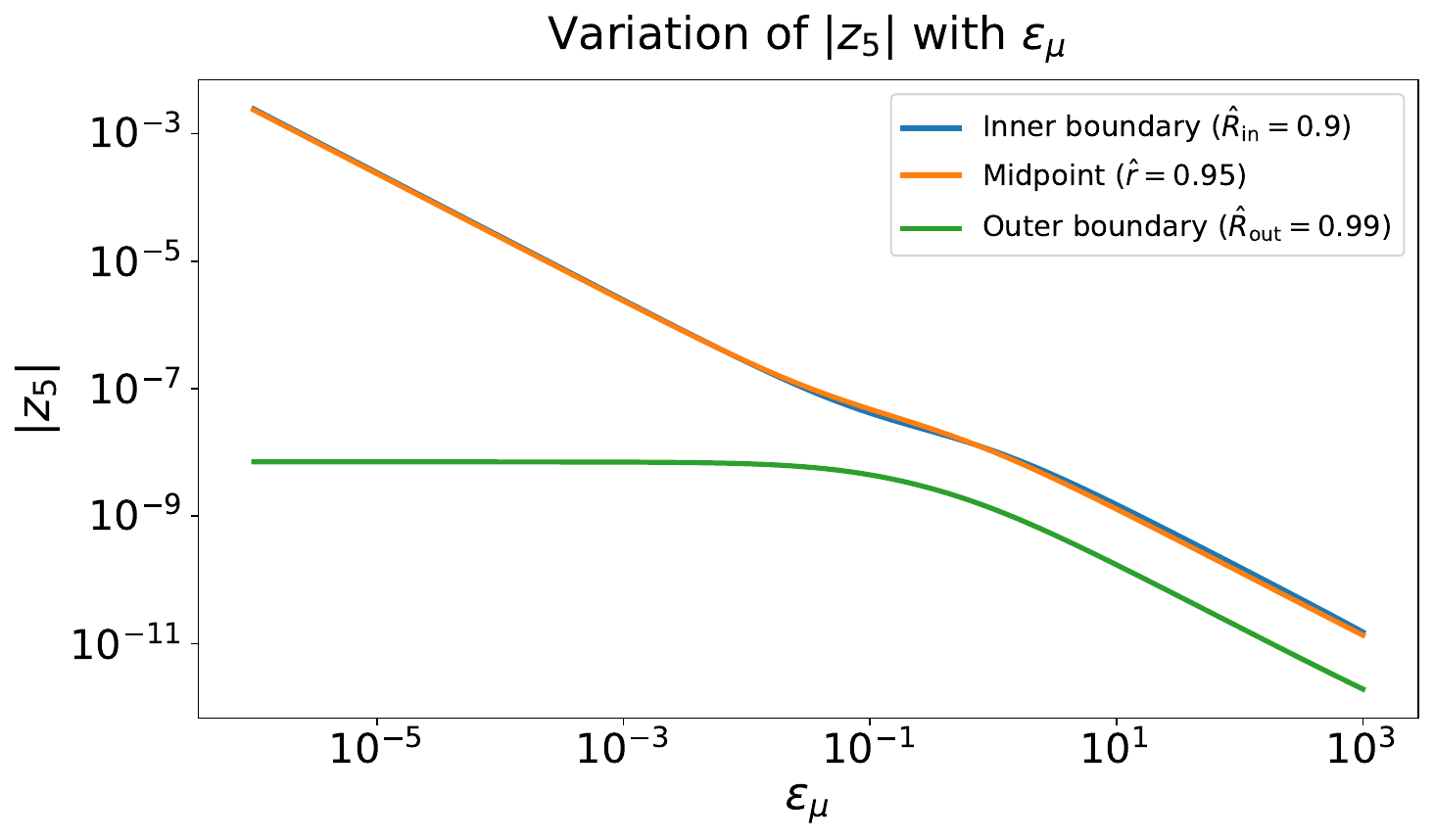}
\\[2mm]
(e)
\end{minipage}

\caption{This figure shows the variation of \( z_1 \), \( z_2 \), \( z_3 \), \( z_4 \), and \( z_5 \) with the dimensionless shear parameter \( \varepsilon_{\mu} \) at the inner crust boundary (blue), midpoint of the crust (orange) and outer crust boundary (green), shown in panels (a), (b), (c), (d), and (e), respectively. In panel (c), the blue and orange curves lie behind the green curve and are therefore not visible. Back-of-the-envelope (BOTE) estimates are also shown for \(z_1\) and \(z_3\). Both the horizontal and vertical axes are plotted on logarithmic scales. Throughout, we fix \(\varepsilon_{\Omega} = 10^{-3}\) and \(\varepsilon_{\mathrm{Mag}} = 10^{-7}\), both of which are parameterized by the stellar spin frequency \(f\), which is taken to be \(100\,\mathrm{Hz}\).}
\label{fig:zi_loglog_mag_ann}

\end{figure*}

\begin{figure*}
    \centering
    \includegraphics[width=\linewidth]{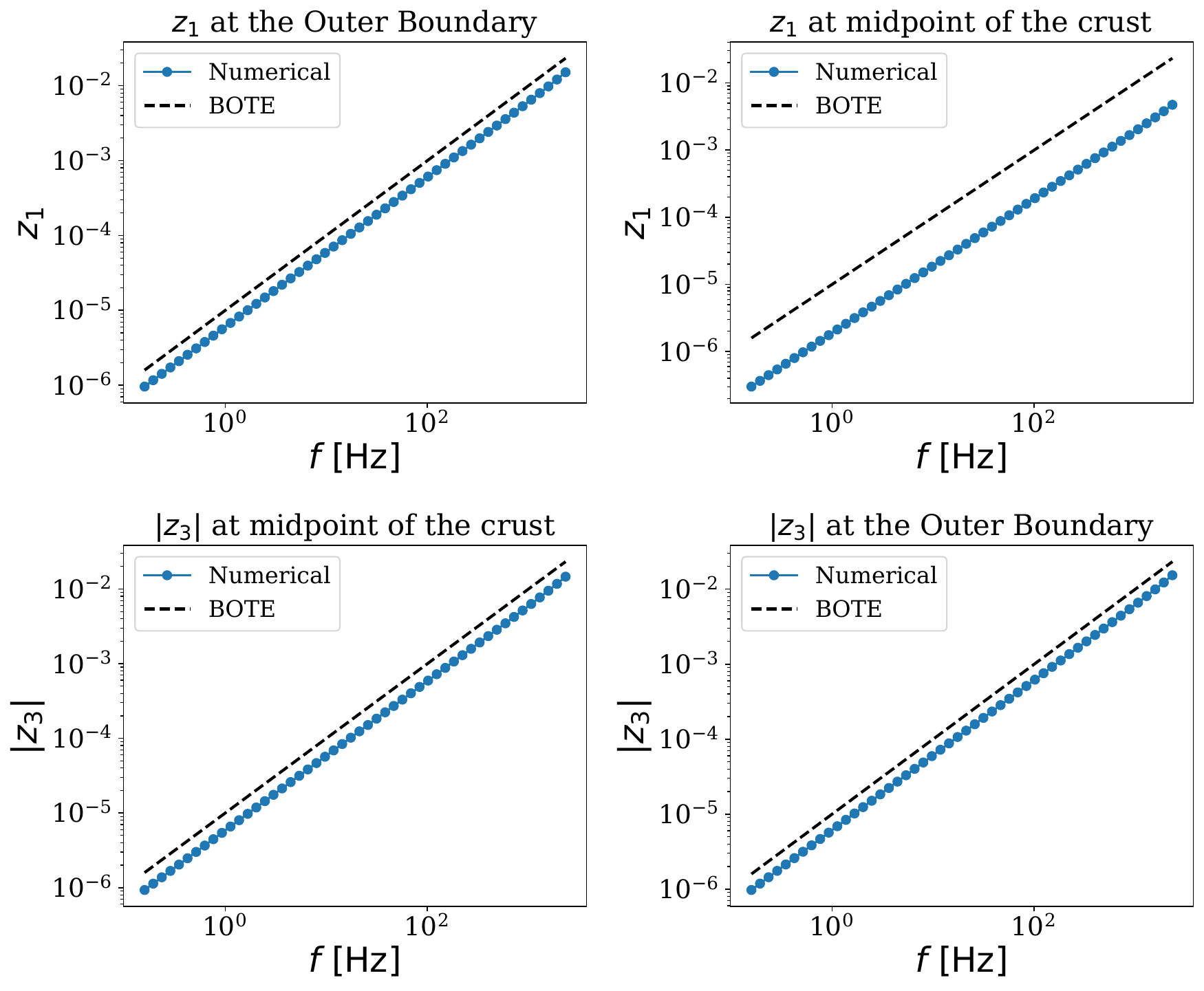}
        
    \caption{This figure shows the variation of \( z_1 \) and \( z_3 \) at the midpoint and the outer boundary of the crust as functions of the spin frequency on a log--log scale. The black dashed line represents the back-of-the-envelope (BOTE) estimate. The top row, panels (right) and (left), corresponds to \( z_1 \) evaluated at the midpoint and the surface, respectively, while the bottom row, panels (left) and (right), shows the corresponding results for \( z_3 \). In all panels, \( \varepsilon_{\mu} = 10^{-5} \). The parameter \( \varepsilon_{\mathrm{Mag}} \) is recomputed for each spin frequency, with the relative critical velocity set to \( 10^{4}\,\mathrm{cm\,s^{-1}} \).}
    \label{fig:z1_3_rotation_ann_log}
\end{figure*}

\begin{figure*}
    \centering    
    \includegraphics[width=\linewidth]{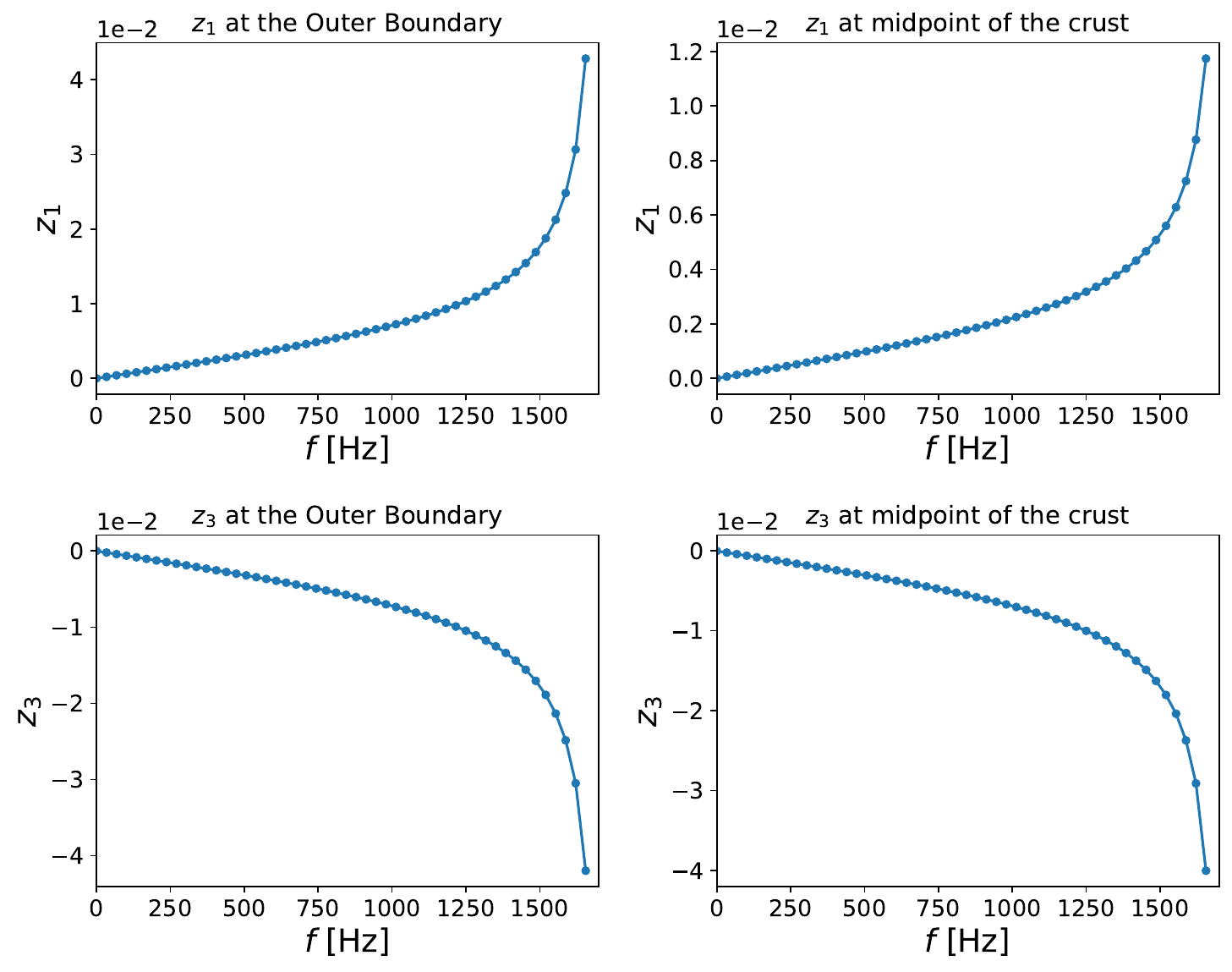}   
    \caption{This figure shows the variation of \( z_1 \) and \( z_3 \) at the midpoint and the outer boundary of the crust as functions of the spin frequency on a linear-linear scale.   The top row displays \( z_1 \), with the left and right panels corresponding to the outer boundary and the midpoint of the crust, respectively, while the bottom row, panels (left) and (right), shows the corresponding results for \( z_3 \). In all panels, \( \varepsilon_{\mu} = 10^{-5} \). The parameter \( \varepsilon_{\mathrm{Mag}} \) is recomputed for each spin frequency, with the relative critical velocity set to \( 10^{4}\,\mathrm{cm\,s^{-1}} \).}
    \label{fig:z1_3_rotation_ann}
\end{figure*}

\begin{figure}
    \centering
    \includegraphics[width=\columnwidth]{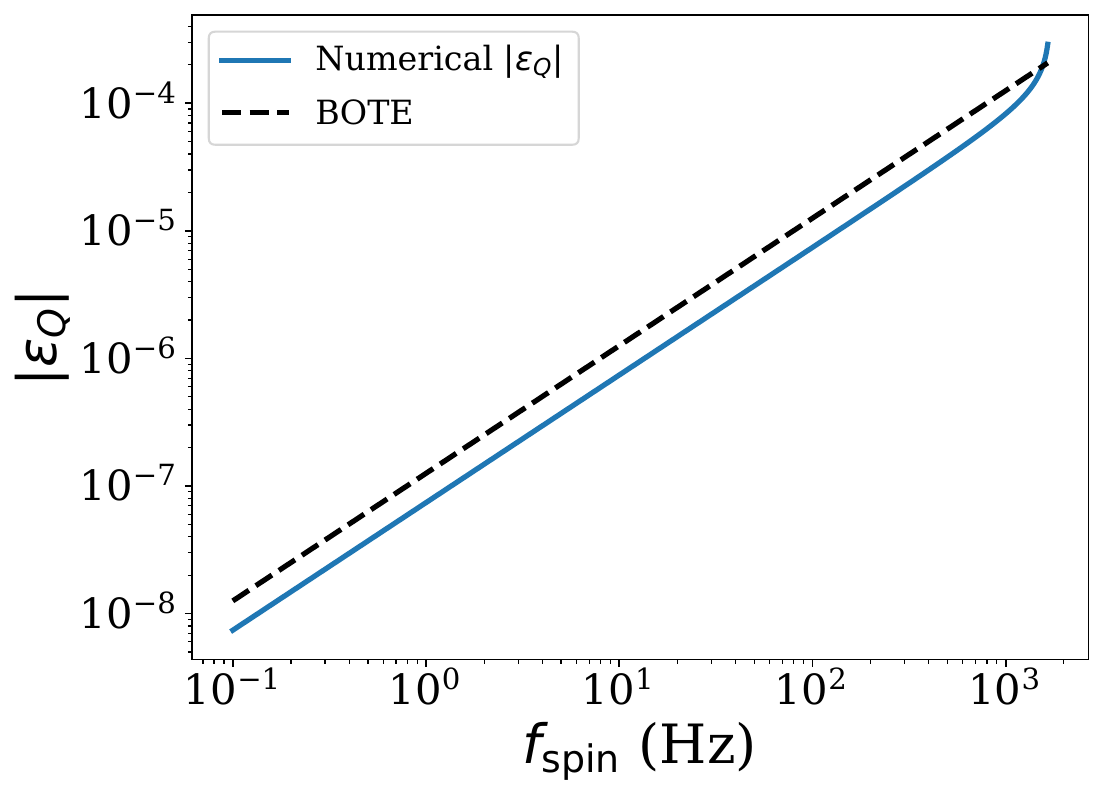}
   \caption{Log--log plot of the absolute quadrupole ellipticity, \(|\varepsilon_Q|\), as a function of the stellar spin frequency \(f\), for \(\varepsilon_{\mu}=10^{-5}\).  We set \(\delta v_{\mathrm{n}} = 10^4\,\mathrm{cm\,s^{-1}}\).}
    \label{fig:logquad}
\end{figure}

\begin{figure}
    \centering
    \includegraphics[width=\columnwidth]{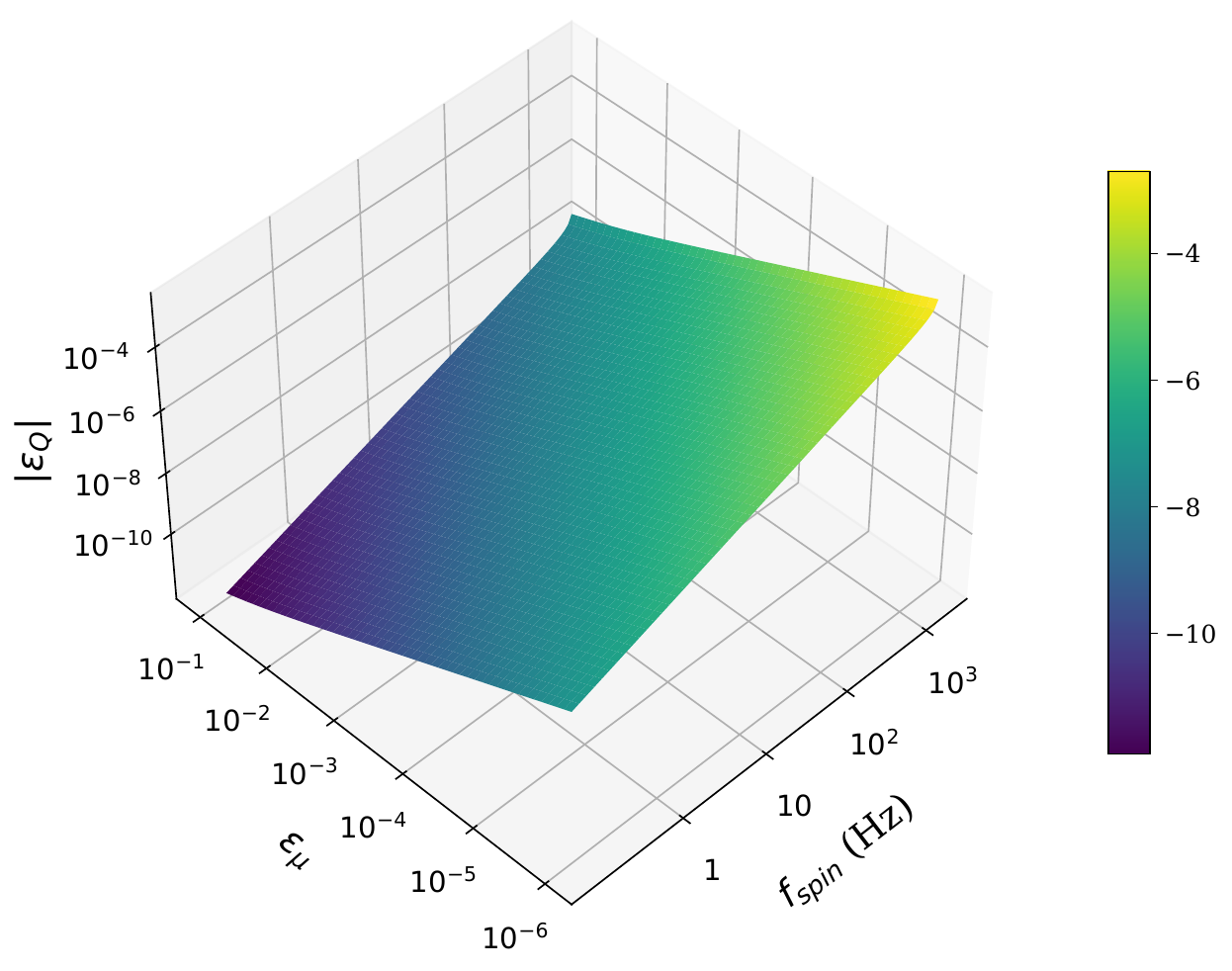}
    \caption{Log--log three-dimensional surface plot of the absolute quadrupole ellipticity, $|\varepsilon_Q|$, as a function of the stellar spin frequency $f$ and the elastic modulus parameter $\varepsilon_\mu$. The parameters, $\varepsilon_{\mathrm{\mathrm{Mag}}}$ and $\varepsilon_{\Omega}$, are scaled with the spin frequency according to the frequency-dependent relations. The colour scale represents $\log_{10}|\varepsilon_Q|$. The figure illustrates the increase in the quadrupole deformation with increasing spin frequency and decreasing values of $\varepsilon_\mu$. We set \(\delta v_{\mathrm{n}} = 10^4\,\mathrm{cm\,s^{-1}}\).}
    \label{quadellip}
\end{figure}

To investigate the dependence of the solutions on the crustal elasticity, in  Figures~\ref{fig:zi_loglog_mag_ann}(a)--\ref{fig:zi_loglog_mag_ann}(e) we show plots of \( z_1 \), \( z_2 \), \( z_3 \), \( z_4 \), and \( z_5 \) as functions of \( \varepsilon_{\mu} \).  We  plot the variables at three representative locations within the crust: the inner boundary, the midpoint of the crust, and the outer boundary.  The quantity \( z_4 \) is plotted only at the midpoint of the crust, since it vanishes at both boundaries due to the imposed boundary conditions. We remind the reader that we expect $\varepsilon_{\mu} = 10^{-5}$ in real neutron stars, with values a few orders of magnitude higher possible for exotic stars with solid cores; we explore a much wider range here to better understand the behaviour of our solutions.  For $z_1$ and $z_3$ we also plot the back-of-the-envelope  estimate of equation~\eqref{z1_est_3}.

In all cases the solutions are smooth functions of $\varepsilon_{\mu}$.  From panels (a) and (c) we see good agreement between the numerical results for $z_1$ and $z_3$ with the BOTE estimate.  Specifically, the numerical results have the predicted scaling with $\varepsilon_{\mu}^{-1}$ over the $\sim$nine orders of magnitude variation plotted.  The numerically computed values of $z_1$ and $z_3$ lie within a factor of $\sim 2$ of the BOTE estimates in all cases, apart from the value of $z_1$ at the midpoint and inner boundary of the crust, where the numerical result is a factor $\sim 5$ smaller than the BOTE estimate.   This is as good an agreement as one can expect given the rough nature of the BOTE estimate.

Next, we examine the variation of \(z_1\) and \(z_3\) at the midpoint of the crust and at the outer boundary, as functions of the stellar spin frequency. The results are presented in Fig.~\ref{fig:z1_3_rotation_ann_log}. We focus only on \(z_1\) and \(z_3\) because they correspond to the radial and azimuthal displacements of the crust and are the quantities that directly enter the calculation of the quadrupole ellipticity. We fix \( \varepsilon_{\mu} = 10^{-5} \). The parameter \( \varepsilon_{\mathrm{Mag}} \) is recomputed for each spin frequency, with the relative critical velocity set to \( 10^{4}\,\mathrm{cm\,s^{-1}} \). We also plot the BOTE estimate of equation~\eqref{z1_est_final}.

Over most of the explored range, \( z_1 \) and \( z_3 \)  lie within a factor of 5 or so of the BOTE, and show the predicted linear scaling with spin frequency at all but the highest spin rates.

To better exhibit the behaviour of  \( z_1 \) and \( z_3 \) at high spin frequencies, we present the plots on a linear scale, as shown in Fig.~\ref{fig:z1_3_rotation_ann}. Significant deviations from the linear trend emerge as the star approaches the break-up spin frequency, with the numerical values increasing more rapidly than linearly.  Since we do not employ a slow-rotation approximation, this nonlinear behavior is captured self-consistently within the model and is to be taken seriously. It in fact has a natural interpretation: close to the Keplerian break-up frequency, the centrifugal force becomes comparable to gravity, substantially reducing the effective restoring force.

The most important quantity of interest in this work is the mountain size, as characterised by the quadrupole ellipticity. Using equation~\eqref{eq:quad_final}, derived in Section~\ref{quad}, we calculate the quadrupole ellipticity from the numerical solutions. Fig.~\ref{fig:logquad} shows a log--log plot of the absolute value of the quadrupole ellipticity, $|\epsilon_Q|$, as a function of the stellar spin frequency, $f$, for a fixed shear parameter, $\varepsilon_\mu=10^{-5}$. Over the frequency range considered, the quadrupole ellipticity exhibits an approximately linear dependence on $f$.

For comparison, we also include a back-of-the-envelope estimate based on the quadrupole ellipticity defined in equation~\eqref{eq:quad_final}. This estimate is given by
\begin{equation}
    \epsilon_Q^{\mathrm{BOTE}}
    \simeq
    \textcolor{black}{\frac{1}{10}}\frac{\Delta R}{R}\,z_1,
\end{equation}
where $\Delta R$ is the crustal thickness and $z_1$ is evaluated using the BOTE estimate given in equation~\eqref{z1_est_final}.  \textcolor{black}{The factor of $1/10$ accounts for the charged-component fraction of the total density, $\rho_{\mathrm p}/\rho = 1/10$, for the adopted neutron-to-charged-component density ratio of $9:1$.} The factor $\Delta R/R$ accounts approximately for the fact that the deformation is supported only by the crust, whereas the moment of inertia entering the definition of the ellipticity (equation~\eqref{eq:quad_final}) receives contributions from the entire star. This scaling can also be seen by comparison of the integrals in the numerator and denominator of equation~\eqref{eq:quad_final}, taking into account the different radial ranges of the integrals.  The resulting estimate is shown by the black dashed line in Fig.~\ref{fig:logquad}. The numerical result and the back-of-the-envelope estimate \textcolor{black}{are of the same order of magnitude}. The predicted linearity is maintained over the four order-of-magnitude range plotted, aside from close to break-up, where again the numerical result is seem to increase more steeply with frequency.

Figure~\ref{quadellip} shows a log--log three-dimensional surface plot of the absolute quadrupole ellipticity, $|\epsilon_Q|$, as a function of the stellar spin frequency $f$ and the shear parameter $\varepsilon_\mu$. The surface is approximately planar over most of the parameter space, and is well described by the fitting formula
\begin{equation}
\label{eq:epsilon_fit}
|\epsilon_Q| \simeq 8.6\times10^{-8} \left( \frac{\delta v_\mathrm{n}}{10^{4}\,\mathrm{cm\,s^{-1}}} \right) \left( \frac{f}{1\,\mathrm{Hz}} \right) \left( \frac{\varepsilon_\mu}{10^{-5}} \right)^{-1}.
\end{equation}
This is our main result. The scalings with $\delta v_\mathrm{n}$, $f$ and $\varepsilon_\mu$ were all anticipated correctly by the BOTE analysis of Section \ref{BOTE}, while the pre-factor has been fit to actual numerical results. As described previously, the three scalings have a simple interpretation. The deforming Magnus force is proportional to both $\delta v_\mathrm{n}$ and $f$, while the ellipticity decreases with increasing $\varepsilon_\mu$, since a larger shear parameter corresponds to a stiffer crust that is more resistant to deformation.  
\textcolor{black}{Equation~\eqref{eq:epsilon_fit} is subject to several uncertainties. The most important limitations arise from the cylindrical geometry and the simplified polytropic equation of state. Hence, the fitting formula should be regarded as an order-of-magnitude estimate within our model rather than a precise prediction for a spherical neutron star. A more realistic calculation would require a spherical stellar model with a realistic EOS.}

%{\bf [DIJ: I'm thinking maybe we should cut this]}
%To verify the correctness of the numerical code, we consider the response of the star to a tidal force, for which the qualitative behaviour of the solutions is well understood. We replace the two-fluid model with a single-component star and use a tidal force in place of the Magnus force, while keeping all other assumptions unchanged. The resulting solutions are smooth and well behaved, and $z_1$ and $z_3$ approach a plateau in the fluid limit, as expected for tidally induced deformations. This provides an independent check of the correctness and stability of the numerical scheme.

\section{Crust breaking criteria}
\label{crust_breaking}

All of our perturbative results, including the crustal displacements and mass quadrupole, are (exactly) linear in the magnitude $\delta v_{\rm n}$ of the imposed neutron superfluid velocity.  In our parameterised results we have scaled this in units of $10^4$\,cm s$^{-1}$, and in our plots (where relevant) have set $\delta v_{\rm n} = 10^4$\,cm s$^{-1}$, motivated by estimates of the critical velocity for unpinning, as per the discussion of Section \ref{BOTE}.  However, there is a second mode of failure that must be taken into account: fracture of the crust itself. 

To explore this issue, we follow \cite{Ushomirsky_2002} and consider the von Mises yielding criterion,
\begin{equation}
    \bar{\sigma} > \bar{\sigma}_{\mathrm{break}},
\end{equation}
which states that the crust yields when the von Mises strain, $\bar{\sigma}$, exceeds the crustal breaking strain. We adopt
$\bar{\sigma}_{\mathrm{break}}=0.1$, motivated by the large breaking strains obtained in the molecular-dynamics simulations of \citet{hk_09}.

The von Mises strain is defined as
\begin{equation}
\label{von_mises}
    \bar{\sigma}^{2}
    =
    \frac{1}{2}\sigma^{ab}\sigma_{ab},
\end{equation}
where $\sigma_{ab}$ is the dimensionless physical shear-strain tensor defined by
\begin{equation}
    \sigma_{ab}
    =
    \frac{\operatorname{Re}\!\left(\delta t_{ab}\right)}{\mu},
\end{equation}
and $\mu$ is the shear modulus. The perturbed shear-stress tensor, $\delta t_{ab}$, is given by
\begin{equation}
\begin{aligned}
    \delta t_{ab} = g_{ab}\delta t_{rr}e^{im\phi}
    + w_{ab}\,2\mu\frac{\xi_r}{r}e^{im\phi}
    - e_{ab}\,2\mu\frac{d\xi_r}{dr}e^{im\phi}
    \\
    + f_{ab}\delta t_{r\phi}
    + \Lambda_{ab}\,2\mu\frac{\xi_\perp}{r},
\end{aligned}
\label{eq:stress_general}
\end{equation}
where
\begin{equation}
    \delta t_{rr}  = \mu\left( \frac{4}{3}\frac{d\xi_r}{dr} -\frac{2}{3}\frac{\xi_r}{r} +\frac{2}{3}m^2\frac{\xi_\perp}{r} \right).
    \label{eq:tau_rr_derivation}
\end{equation}

The perturbed shear-stress tensor in equation~\eqref{eq:stress_general} can be obtained directly from
equation~\eqref{per_stress} by removing the isotropic stress contribution.
Here, $\delta t_{r\phi}$ is the same quantity as that given in
equation~\eqref{tau_rperp}. The tensors $g_{ab}$, $w_{ab}$, $e_{ab}$,
$f_{ab}$, and $\Lambda_{ab}$ are defined in
Appendix~\ref{app_pert_eom}.

We substitute the real part of the perturbed shear-stress tensor, equation~\eqref{eq:stress_general}, into the von Mises strain expression, equation~\eqref{von_mises}. We then set $m=2$ and use the definitions of $z_1$ and $z_3$.  \textcolor{black}{Expressing the resulting
equation in terms of the dimensionless radial coordinate
$\hat{r}$,} we obtain
\begin{equation}
\begin{aligned}
    \bar{\sigma}^2(\textcolor{black}{\hat{r}},\phi) = \frac{4}{3} \Bigg[ 
    \textcolor{black}{\hat{r}}^2\left(\frac{dz_1}{d\textcolor{black}{\hat{r}}}\right)^2 
    + \textcolor{black}{\hat{r}} z_1\frac{dz_1}{d\textcolor{black}{\hat{r}}} 
    + z_1^2  + 4z_3\left( \textcolor{black}{\hat{r}}\frac{dz_1}{d\textcolor{black}{\hat{r}}}-z_1 \right)
    \\
    + 16z_3^2 \Bigg]\cos^2(2\phi) 
    + 4\left( \textcolor{black}{\hat{r}}\frac{dz_3}{d\textcolor{black}{\hat{r}}}+z_1 \right)^2 \sin^2(2\phi).
\end{aligned}
\label{eq:von_mises_z1_z3_m2}
\end{equation}

Evaluating this numerically, we find that the strain is largest at the outer crustal boundary, and at the azimuthal angle $\phi = 0$.  It also peaks at $\phi = \pi/2, \pi, 3\pi/2$, a consequence of the $m=4$ symmetry of equation~\eqref{eq:von_mises_z1_z3_m2}, which itself is a consequence of $\bar{\sigma}^2$ being the square of an $m=2$ quantity.

\begin{figure}
    \centering
    \includegraphics[width=\linewidth]{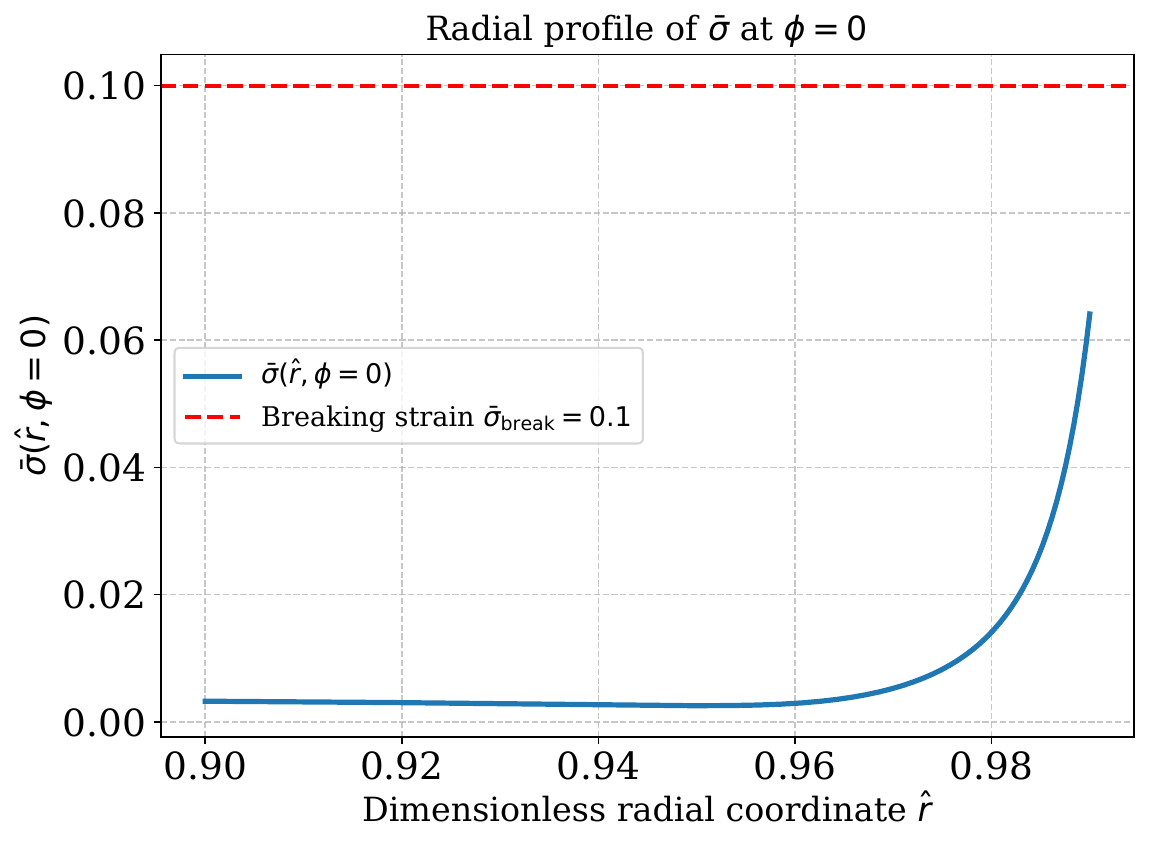}
    \caption{The von Mises strain, $\bar{\sigma}$, at $\phi=0$ across the radial profile of the crust, calculated using the dimensionless parameter values $\varepsilon_{\mu}=10^{-5}$, $\varepsilon_{\Omega}=10^{-3}$, corresponding to $f_{\mathrm{spin}}=100\,\mathrm{Hz}$, and $\varepsilon_{\mathrm{Mag}}=10^{-7}$.}
    \label{fig:radial_profile_sigma}
\end{figure}

To illustrate the radial dependence of the strain, we plot $\bar{\sigma}$ as a function of radius at $\phi=0$ in Fig.~\ref{fig:radial_profile_sigma}, for a star with $\varepsilon_{\mu}=10^{-5}$, $\varepsilon_{\Omega}=10^{-3}$, corresponding to $f_{\mathrm{spin}}=100\,\mathrm{Hz}$, and $\varepsilon_{\mathrm{Mag}}=10^{-7}$. The strain increases rather sharply towards the outer boundary of the crust and reaches its maximum value, $\bar{\sigma}\simeq 0.06$, at $\hat{R}_{\mathrm{out}}=0.99$.

%The azimuthal dependence is evident from equation~\eqref{eq:von_mises_z1_z3_m2}. The strain exhibits a four-fold angular symmetry, with the pattern repeating every $\pi/2$, as expected for an $m=2$ perturbation. For the present solution, the four equivalent maxima, located at the azimuthal positions specified in equation~\eqref{phi_max}, are aligned with the principal axes.

%For the fiducial dimensionless parameter values
%$\varepsilon_{\mu}=10^{-5}$,
%$\varepsilon_{\Omega}=10^{-3}$, and
%$\varepsilon_{\mathrm{Mag}}=10^{-7}$, we evaluate
%$\bar{\sigma}(r,\phi)$ numerically at each grid point within the crust and determine its global maximum. %We find
%\begin{equation}
%    \bar{\sigma}_{\max} \simeq 0.06,
%\end{equation}
%which occurs at the outer boundary of the crust, $r=0.99$, and at
%\begin{equation}
%\label{phi_max}
%    \phi = n\frac{\pi}{2}, \qquad n=0,1,2,3.
%\end{equation}
%Since this value is smaller than the crustal breaking strain, $\bar{\sigma}_{\mathrm{break}}=0.1$, the %crust does not yield before vortex unpinning for the fiducial model.

\begin{figure}
    \centering
    \includegraphics[width=\linewidth]{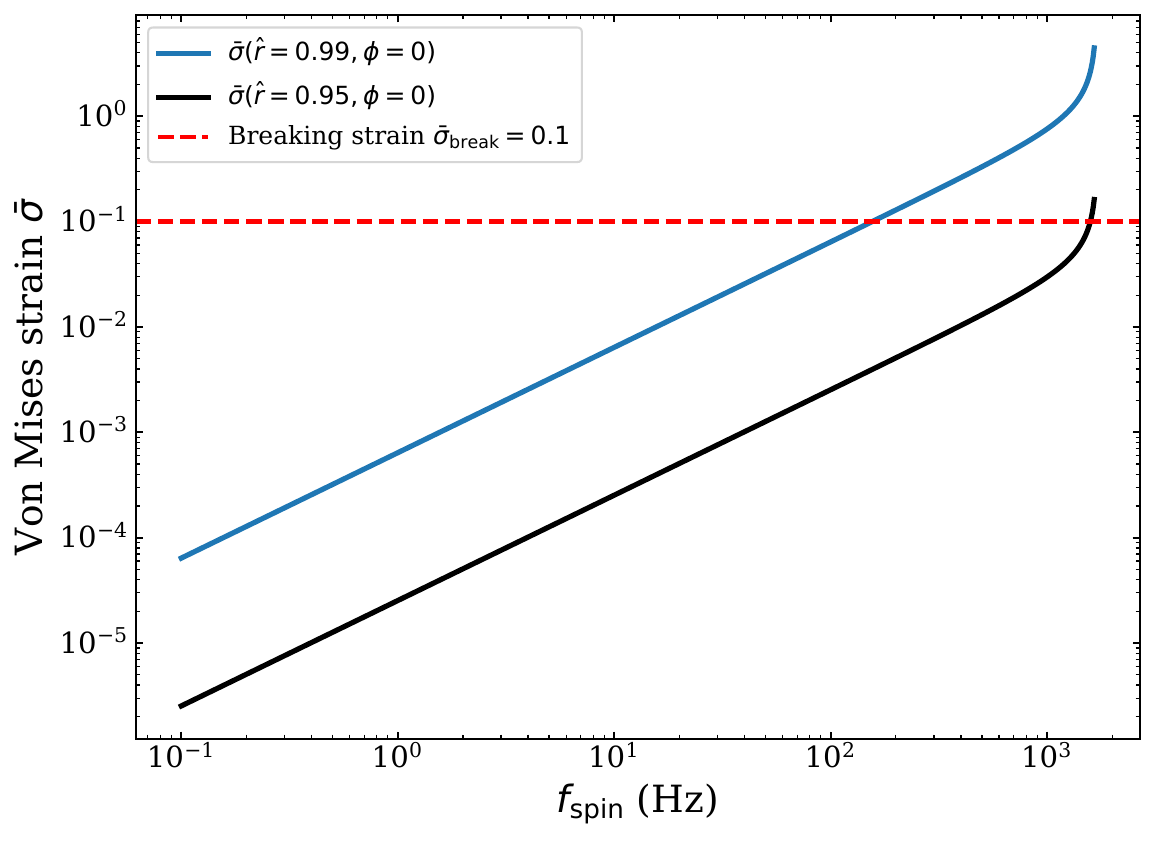}
    \caption{Log-log plot of the von Mises strain, $\bar{\sigma}$, as a function of spin frequency at $\phi=0$, evaluated at the outer crustal boundary, $\hat{R}_{\mathrm{out}}=0.99$, and at the midpoint of the crust, $\hat{r}=0.95$. The blue solid curve corresponds to $\hat{R}_{\mathrm{out}}=0.99$, while the black solid curve corresponds to $\hat{r}=0.95$. The red dashed line marks the adopted crustal breaking strain, $\bar{\sigma}_{\mathrm{break}}=0.1$. The dimensionless shear-modulus parameter is fixed at $\varepsilon_{\mu}=10^{-5}$.}
    \label{fig:sigma_fre}
\end{figure}

We now examine how the strain depends on the spin frequency of the star. We evaluate $\bar{\sigma}$ at $\hat{R}_{\mathrm{out}}=0.99$ and $\phi=0$, corresponding to one of the locations at which the global maximum occurs, over a range of spin frequencies. The result is shown by the blue curve in Fig.~\ref{fig:sigma_fre}. We find that the von Mises strain reaches the breaking threshold at
\begin{equation}
    f_{\mathrm{spin}} \simeq \textcolor{black}{155}~\mathrm{Hz}.
\end{equation}

Thus, within the assumptions of the present model, the crust is expected to yield before vortex unpinning for spin frequencies above approximately $\textcolor{black}{155}~\mathrm{Hz}$. \textcolor{black}{The corresponding quadrupole ellipticity is $\epsilon_Q \approx 10^{-5}$.}

For comparison, the black curve in Fig.~\ref{fig:sigma_fre} shows the strain at the midpoint of the crust, $\hat{r}=0.95$. At this location, the strain remains below the breaking threshold over most of the relevant frequency range and approaches the threshold only near the stellar break-up frequency. This suggests that a substantial fraction of the deeper crust may remain unfractured even after the outermost layers begin to yield, and may therefore continue to support part of the deformation.

However, the onset of failure in the outer crust does not necessarily remain confined to that region. A crack initiated in the low-density outer layers could, in principle, propagate into deeper parts of the crust even where the local strain remains below the nominal breaking threshold. Modelling such fracture propagation  lies outside the scope of the present work.  

Over most of the frequency range, $z_1$ and $z_3$ scale approximately linearly with the spin frequency, and consequently $\bar{\sigma}$ also shows an approximately linear dependence on $f_{\mathrm{spin}}$. Deviations from this behaviour appear only as the stellar spin approaches the break-up regime.

\section{Summary and conclusion}
\label{summary_ponclusion}

We have constructed a simple model for the formation of a Magnus mountain in a two-component cylindrical star. One component represents the neutron superfluid, while the other represents the charged component, which possesses a non-zero shear modulus in the annular crustal region.  We introduced by hand an $m=2$ perturbation to the neutron fluid velocity field.  The non-axisymmetric Magnus force acting on pinned superfluid vortices then served as the source of a deformation. 

We found that the current quadrupole associated with our $m=2$ velocity perturbation was exactly zero, so there is no current quadrupole radiation in our model.
%One can calculate the current multipole associated with the $m=2$ velocity perturbation associated with our chosen velocity field (given in equation (\ref{per_vel_ann})). However, it unexpectedly turns out to be zero, even though a non-zero current multipole is generally expected for an $m=2$ velocity perturbation. 
A more realistic perturbed velocity field, together with a more realistic (i.e. spherical) stellar geometry, may instead yield a non-zero current quadrupole moment.

Related to this, in \cite{Melatos_2015}, the authors considered an infinitely long cylindrical star and found a non-zero current multipole arising from non-axisymmetric pinning of vortices in the crust. However, rather than integrating over the cylindrical domain, they evaluated the current-multipole integral over a spherical volume. 
%defined by \( |z| \le (R^{*2}-R^2)^{1/2} \), where \( R \) varies from \(0\) to \(R^*\), the stellar radius.  
%As a result, the velocity field may no longer satisfy the appropriate boundary conditions on the integration surface. 
A  self-consistent treatment would require both the stellar model and the multipole integral to be formulated within the same stellar volume. Interestingly, \cite{universe8120619} considered a spherical star and obtained a vanishing current multipole.

Despite the vanishing of the current multipole, we obtained a non-zero mass quadrupole moment. \textcolor{black}{We found that the neutron component has vanishing radial displacement at the stellar surface. Whereas the charged component develops a non-zero elastic displacement field.} 

The numerical solutions for the (dimensionless) radial and angular crustal displacement variables \(z_1\) and \(z_3\) were smooth across the radial profile of the star. We examined the variation of \(z_1\) and \(z_3\) with respect to the shear parameter. As expected, increasing the shear modulus makes the crust more resistant to deformation, resulting in a monotonic decrease in the displacement amplitude.

We then examined the variation of \(z_1\) and \(z_3\) with the stellar spin frequency, up to the break-up frequency. Both variables remain approximately linear over most of the spin-frequency range and deviate from linearity only near the break-up frequency. This deviation occurs because, as the star approaches break-up, the centrifugal force increasingly balances gravity, thereby reducing the effective restoring force.

In this work, we are primarily interested in the large-scale deformation produced by the Magnus force acting on pinned superfluid vortices, which serves as the source of gravitational-wave (GW) emission. For this we estimated the stellar deformation provided by the quadrupole ellipticity, as defined in equation~\eqref{eq:quad_final}. From our numerical results, we derive the scaling relation for the quadrupole ellipticity given in equation~\eqref{eq:epsilon_fit}. 

%This scaling relation is the main result of this paper.

We also investigate whether the crust yields before vortex unpinning occurs. For this, we apply the von Mises yielding criterion and adopt a crustal breaking strain of
$\bar{\sigma}_{\mathrm{break}}=0.1$. At a spin frequency of $100\,\mathrm{Hz}$, the maximum von Mises strain remains below this threshold, indicating that vortex unpinning occurs before crustal failure. As the spin frequency increases, however, the maximum von Mises strain in the crust also increases and reaches the breaking threshold at $f_{\mathrm{spin}} \simeq \textcolor{black}{155}\,\mathrm{Hz}$. Thus, within the assumptions of our model, the crust is expected to yield before vortex unpinning for spin frequencies above approximately $\textcolor{black}{155}\,\mathrm{Hz}$. At this threshold frequency, the corresponding quadrupole ellipticity is $\epsilon_Q \approx 10^{-5}$. This therefore represents the largest mountain that can be supported in our model without any part of the crust exceeding the breaking strain threshold.

\textcolor{black}{\textcolor{black}{The maximum ellipticity obtained in our model, $\varepsilon_Q\sim5\times10^{-5}$, is larger than the characteristic values of order $10^{-6}$ found in previous studies of elastic mountains \citep{Ushomirsky_2002, MoralesHorowitz2022}. A direct comparison is complicated by differences in the underlying models. Two main differences that could contribute to this discrepancy are the cylindrical geometry considered here and the use of a two-fluid model. Most previous calculations employ a single-fluid stellar model with an elastic crust, whereas the present calculation uses a two-fluid description with independent neutron-superfluid and charged-component dynamics. The recent spherical Magnus-mountain calculation of \citet{Giliberti2026} gives ellipticities about an order of magnitude smaller than ours. However, the two calculations differ in their underlying physical assumptions: \citet{Giliberti2026} considers a one-component spherical model with realistic equations of state and a different treatment of the crustal response, whereas the present calculation uses a two-component cylindrical model. The author argues that the discrepancy cannot be attributed to geometry alone and may also arise from the additional relative degree of freedom present in the two-fluid model. A two-fluid Magnus-mountain calculation in spherical geometry will be required to separate these effects and will be an important direction for future work.}}

Our results further indicate that, although the low-density outer crust begins to yield at approximately $\textcolor{black}{155}\,\mathrm{Hz}$, most of the deeper crust remains below the breaking strain threshold, $ \bar{\sigma}_{\mathrm{break}} =0.1$, up to frequencies close to the stellar break-up limit. The deeper elastic layers may therefore continue to support the mountain after yielding begins in the outer crust. Under this assumption, the largest mountain predicted by our model has a quadrupole ellipticity of
\(\epsilon_Q \approx 5\times10^{-5}\) at a stellar spin frequency of \(700\,\mathrm{Hz}\), comparable to the spin frequencies of the fastest observed neutron stars. However, because modelling fracture propagation lies beyond the scope of the present work, we cannot determine with certainty whether such a mountain can be sustained following the onset of crustal failure.
 
Several open questions remain to be addressed in future analyses, along with potential modifications and enhancements to the proposed setup. Some of these are outlined below:

\begin{enumerate}

    \item In this study, we specified a $m=2$ perturbed neutron superfluid velocity field by hand. A key direction for future work is to evaluate this velocity field from first principles.  This may come from the spontaneous generation of asymmetry in the dynamics of interacting vortices, as is suggested by the works of \citet{Melatos_2015}, \citet{cheunchitra2024persistent} and \citet{viswanathan2026}. However, such  numerical simulations can model only a tiny fraction of the vortices present in a realistic neutron star. For a neutron star rotating at \(\sim100\,\mathrm{Hz}\), the expected number of vortices is of order \(10^{17}\!-\!10^{18}\), whereas existing simulations typically evolve only \(10^2\!-\!10^5\) vortices because of computational limitations. Consequently, the results must be extrapolated over many orders of magnitude, and this limitation should be kept in mind when interpreting their astrophysical implications.
    
    Another approach would be to consider the microphysical interaction of the vortices with the pinning potentials associated with flux tubes and crustal nuclei.  Some ideas in this direction have been explored by \citet{Sidery2009TheEO}.
    
    \item The analysis should be extended to a realistic spherical configuration of the star. The infinite cylindrical configuration was chosen for simplicity, as it aligns with the natural geometry of the vortices, with no perturbations along the $z$-axis. However, in a spherical geometry, this assumption no longer holds. Moreover, it is necessary to consider a spherical configuration to determine whether it may give rise to a non-zero current multipole.

    \item Once the model is extended to the spherical configuration, a further refinement could involve conducting the analysis within the framework of General Relativity (GR) instead of Newtonian mechanics. 
    
\end{enumerate}

Finally, we point out that, even taking into account the limitations on Magnus mountain size imposed by the crust's finite breaking strain, the ellipticities we find for our cylindrical stars, as parameterised by equation~\eqref{eq:epsilon_fit},  are large.  Such values, if realised in realistic neutron stars, would lie well within the reach of even current generation gravitational wave detectors, and in many cases have already been ruled out by non-detections \citep{jr_24}. \textcolor{black}{Whether realistic neutron stars can sustain such large
ellipticities remains an open question. Our cylindrical calculation serves only as a proof of concept for the formation of Magnus mountains through the two-fluid mechanism. To place meaningful constraints on their ellipticity, we need to extend the two-fluid calculation to spherical stars.} This clearly motivates more detailed work on Magnus mountains.

%%%%%%%%%%%%%%%%%%%%%%%%%%%%%%%%%%%%%%%%%%%%%%%%%%%%%%%%%%%%%%%%%%5

\section*{Acknowledgements}

YG acknowledges support from the Engineering and Physical Sciences Research Council (EPSRC) through grant No.~EP/W524013/1. DIJ acknowledges support from the Science and Technology Facilities Council (STFC) through grants Nos.~ST/R00045X/1 and APP46132.

The authors are grateful to Toby Wood for helpful discussions on the boundary conditions employed in the numerical calculations. They also thank Professor Carsten Gundlach for useful discussions on regularity conditions in cylindrical coordinates and Professor Ian Hawke for discussions on numerical methods. The authors further thank Kostas Glampedakis for his contributions to this area through earlier collaborative work with DIJ.

%%%%%%%%%%%%%%%%%%%%%%%%%%%%%%%%%%%%%%%%%%%%%%%%%%
\section*{Data Availability}
The data underlying this article are available from the corresponding author upon reasonable request.



\bibliographystyle{mnras}
\bibliography{example} % if your bibtex file is called example.bib

%\begin{thebibliography}{99}
%\end{thebibliography}

%%%%%%%%%%%%%%%%%%%%%%%%%%%%%%%%%%%%%%%%%%%%%%%%%%

%%%%%%%%%%%%%%%%% APPENDICES %%%%%%%%%%%%%%%%%%%%%

\appendix

\section{Estimate of $K_{\mathrm{\lowercase{n}}}$ and $K_{\mathrm{\lowercase{p}}}$}
\label{Back_env}

To set the values of $K_\mathrm{n}$ and $K_\mathrm{p}$, we fix the radius of the star as $R = 10 \, \text{km}$. Using the ratio of the mass per unit length of the neutron superfluid,
$\lambda_{\mathrm{n}}$, to that of the charged component,
$\lambda_{\mathrm{p}}$, we write

\begin{equation}
    \frac{\lambda_\mathrm{n}}{\lambda_\mathrm{p}} = \frac{\int_0^R \rho_{\mathrm{n}}r \, dr \, d\phi}{\int_0^R \rho_{\mathrm{p}}r \, dr \, d\phi} = \frac{2 \pi \frac{m_\mathrm{n}^2}{K_\mathrm{n}} \int_0^R  \Tilde{ \mu_0} r \, dr}{2 \pi \frac{m_\mathrm{p}^2}{K_\mathrm{p}} \int_0^R  \Tilde{ \mu_0} r \, dr} = \frac{m_\mathrm{n}^2 K_\mathrm{p}}{m_\mathrm{p}^2 K_\mathrm{n}}.
\end{equation}

In a typical neutron star, we assume the following ratio of neutron mass density to proton mass density ~\citep{Lattimer:2004pg}:
\begin{equation}
\label{lambda_ratio}
    \frac{\lambda_\mathrm{n}}{\lambda_\mathrm{p}} = 9.
\end{equation}
Using this ratio, we get:
\begin{equation}
     \frac{9}{1} = \frac{m_\mathrm{n}^2 K_\mathrm{p}}{m_\mathrm{p}^2 K_\mathrm{n}}.
\end{equation}

Assuming $m_{\mathrm{n}}\approx m_\mathrm{p}$, this simplifies to:
\begin{equation}
    \frac{K_\mathrm{p}}{ K_\mathrm{n}} = 9.
\end{equation}

Setting $\Omega=0$ in equation~\eqref{eq:mu_final}, the reduced chemical
potential in the non-rotating limit becomes
\begin{equation}
    \Tilde{ \mu_0} = \frac{2 G \lambda}{R \sqrt{A} J_1 (\sqrt{A} R)}J_0 (\sqrt{A} r).
\end{equation}
Imposing the boundary condition that the reduced chemical potential
vanishes at the stellar surface, i.e.
$\tilde{\mu}_0(R)=0$, gives
\begin{equation}
    \frac{2 G \lambda}{R \sqrt{A} J_1 (\sqrt{A} R)}J_0 (\sqrt{A} R) = 0.
\end{equation}

Since the prefactor is non-zero, we require:
\begin{equation}
    J_0 (\sqrt{A} R) = 0,
\end{equation}
where
\begin{equation}
   A =  4 \pi G \left( \frac{m^2_\mathrm{n}}{K_\mathrm{n}} + \frac{m^2_\mathrm{p}}{K_\mathrm{p}} \right).
\end{equation}

Using the first root of the Bessel function \citep{mathworld:BesselZero}:
\begin{equation}
    \sqrt{A} R \approx 2.4048,
\end{equation}
and squaring both sides gives:
\begin{equation}
    A R^2 \approx 5.78.
\end{equation}

Substituting $A$:
\begin{equation}
    4 \pi G \left( \frac{m^2_\mathrm{n}}{K_\mathrm{n}} + \frac{m^2_\mathrm{p}}{K_\mathrm{p}} \right) R^2 \approx 5.78.
\end{equation}

Given $R = 10 \, \text{km} = 10^6 \, \text{cm}$, and using $K_{\mathrm{p}}= 9K_\mathrm{n}$, we get:
\begin{equation}
    4 \pi G \left( \frac{m^2_\mathrm{n}}{K_\mathrm{n}} + \frac{m^2_\mathrm{p}}{9 K_\mathrm{n}} \right) 10^{12} \approx 5.78.
\end{equation}
With $m_{\mathrm{n}}\approx m_{\mathrm{p}}\approx 1.67 \times 10^{-24} \, \text{g}$, we obtain:
\begin{equation}
\label{k_n_def}
    K_{\mathrm{n}}\approx 4.49 \times 10^{-43} \, \text{g} \cdot \text{cm}^5 \cdot \text{s}^{-2},
\end{equation}
\begin{equation}
\label{k_p_def}
    K_{\mathrm{p}}\approx 4.04 \times 10^{-42} \, \text{g} \cdot \text{cm}^5 \cdot \text{s}^{-2}.
\end{equation}

These values of $K_\mathrm{n}$ and $K_\mathrm{p}$ are then used to plot the density profiles $\rho_\mathrm{n}$ and $\rho_\mathrm{p}$ of the background star, shown in Fig.~\ref{density_purve}.

\textcolor{black}{The charged-component pressure is defined in equation~\eqref{psi_not}. For completeness, we similarly define the neutron pressure as
\begin{equation}
\Psi_{\mathrm n} = \frac{m_{\mathrm n}^2}{2K_{\mathrm n}} \tilde{\mu}_{\mathrm n}^{\,2},
\end{equation}
For the fiducial model, the central pressures are
\begin{equation}
\Psi_0 = \Psi_{\mathrm p,0} \simeq 3.9\times10^{34}\,\mathrm{dyn\,cm^{-2}},
\label{eq:central_pressure_p}
\end{equation}}\textcolor{black}{
\begin{equation}
\Psi_{\mathrm n,0} \simeq 3.5\times10^{35}\,\mathrm{dyn\,cm^{-2}},
\label{eq:central_pressure_n}
\end{equation}}\textcolor{black}{
\begin{equation}
\Psi_{\mathrm{tot},0} = \Psi_{\mathrm p,0} +\Psi_{\mathrm n,0} \simeq 3.9\times10^{35}\,\mathrm{dyn\,cm^{-2}}.
\label{eq:central_pressure_tot}
\end{equation}}

\section{perturbed stress tensor in cylindrical harmonics}
\label{app_pert_eom}

In \cite{Ushomirsky_2002}, the authors expressed the perturbed stress tensor, given in equation~\eqref{eq:pert_stress_tensor_pom}, in terms of the perturbed tractions \(\delta \tau_{rr}\) and \(\delta \tau_{r\perp}\) in spherical coordinates. This formulation facilitates the application of boundary conditions at the surface. Inspired by this approach, we write a general form of the perturbed stress tensor in cylindrical coordinates:
\begin{multline}
\delta \tau_{ab} = g_{ab} \delta \tau_{rr} C_{mk} + w_{ab} 2 \mu \frac{\xi_r}{r} C_{mk} - e_{ab} 2 \mu \frac{d \xi_r}{dr} C_{mk} \\ + f_{ab} \delta \tau'_{r\phi} + \tilde{f}_{ab} \delta \tau'_{rz} + \Lambda_{ab} 2 \mu \frac{\xi_\perp}{r},
\end{multline}
where 
\begin{equation}
C_{mk} = e^{im\phi}e^{ikz}.
\end{equation}

The individual components are given by
\begin{align}
\label{tau_rr_incomp}
\delta \tau_{rr} &= - \rho_{\mathrm{p}}\delta \tilde{\mu}_{\mathrm{p}}+ \mu \left( \frac{4}{3} \frac{d \xi_r}{dr} - \frac{2}{3} \frac{\xi_r}{r} + \frac{2}{3} m^2 \frac{\xi_\perp}{r} + \frac{2}{3} k^2 \xi_\perp r \right), \\
\delta \tau'_{r\phi} &= \mu \left( \frac{d \xi_\perp}{dr} - \frac{\xi_\perp}{r} + \frac{\xi_r}{r} \right), \\
\delta \tau'_{rz} &= \mu \left( r \frac{d \xi_\perp}{dr} + \xi_\perp + \xi_r \right), \\
f_{ab} &= r (\hat{r}_a \nabla_b C_{mk} + \hat{r}_b \nabla_a C_{mk} - \hat{r}_a \hat{z}_b i k C_{mk} - \hat{r}_b \hat{z}_a i k C_{mk}), \\
\tilde{f}_{ab} &= \hat{r}_a \nabla_b C_{mk} + \hat{r}_b \nabla_a C_{mk} - \frac{\hat{r}_a \hat{\phi}_b i m C_{mk}}{r} - \frac{\hat{r}_b \hat{\phi}_a i m C_{mk}}{r}, \\
\label{lambda_ab}
\Lambda_{ab} &= r^2 \nabla_a \nabla_b C_{mk} + f_{ab}, \\
\label{e_ab}
e_{ab} &= g_{ab} - \hat{r}_a \hat{r}_b, \\
\label{w_ab}
w_{ab} &= e_{ab} - \hat{z}_a \hat{z}_b.
\end{align}

In our specific model, we do not include perturbations along the \(z\)-axis. Therefore, the expressions simplify to:
\begin{multline}
\label{per_stress}
\delta \tau_{ab} = g_{ab} \delta \tau_{rr} e^{im\phi} + w_{ab} 2 \mu \frac{\xi_r}{r} e^{im\phi} - e_{ab} 2 \mu \frac{d \xi_r}{dr} e^{im\phi} \\ + f_{ab} \delta \tau'_{r\phi} + \Lambda_{ab} 2 \mu \frac{\xi_\perp}{r},
\end{multline}
where:
\begin{align}
\label{tau_rr}
\delta \tau_{rr} &= - \rho_{\mathrm{p}}\delta \tilde{\mu}_{\mathrm{p}}+ \mu \left( \frac{4}{3} \frac{d \xi_r}{dr} - \frac{2}{3} \frac{\xi_r}{r} + \frac{2}{3} m^2 \frac{\xi_\perp}{r} \right), \\
\label{tau_rperp}
\delta \tau'_{r\phi} &= \mu \left( \frac{d \xi_\perp}{dr} - \frac{\xi_\perp}{r} + \frac{\xi_r}{r} \right), \\
\label{f_ab}
f_{ab} &= r (\hat{r}_a \nabla_b e^{im\phi} + \hat{r}_b \nabla_a e^{im\phi}), \\
\xi^a &= \xi_r e^{im\phi} \hat{r}^a + \xi_\perp r \nabla^a e^{im\phi}.
\end{align}

Finally, substituting the expression for the perturbed stress tensor from equation~\eqref{per_stress} into the perturbed EOM~\eqref{eq:eom_elastic_pomp_3}, we obtain:
\begin{multline}
\label{eom_e1}
\nabla^a \left(
g_{ab} \delta \tau_{rr} e^{im\phi}
+ w_{ab} 2 \mu \frac{\xi_r}{r} e^{im\phi}
- e_{ab} 2 \mu \frac{d \xi_r}{dr} e^{im\phi}
\right. \\
\left.
+ f_{ab} \delta \tau'_{r\phi}
+ \Lambda_{ab} 2 \mu \frac{\xi_\perp}{r}
\right)
+ \delta \rho_{\mathrm{p}}\nabla_b \tilde{\mu}_{\mathrm{p}} e^{im\phi} =
\rho_{\mathrm{n}}2 \Omega \epsilon_{bzd} \delta v_\mathrm{n}^d .
\end{multline}

Inserting (\ref{lambda_ab}), (\ref{e_ab}), (\ref{w_ab}) and (\ref{f_ab}) into (\ref{eom_e1}) gives,
\\
\begin{multline}
\label{eom_e2}
\nabla^a \Big(
g_{ab} \delta \tau_{rr} e^{im\phi}
+ g_{ab} e^{im\phi} 2\mu \left( \frac{\xi_r}{r} - \frac{d\xi_r}{dr} \right)
- \hat{r}_a \hat{r}_b e^{im\phi} 2\mu \left( \frac{\xi_r}{r} - \frac{d\xi_r}{dr} \right) \\
- \hat{z}_a \hat{z}_b 2\mu e^{im\phi} \frac{\xi_r}{r}
+ r(\hat{r}_a \nabla_b e^{im\phi} + \hat{r}_b \nabla_a e^{im\phi})
\left( \delta \tau'_{r\phi} + 2\mu \frac{\xi_\perp}{r} \right) \\
+ r^2 \nabla_a \nabla_b C_{mk}\, 2\mu \frac{\xi_\perp}{r}
\Big)
+ \delta \rho_{\mathrm{p}}\nabla_b \tilde{\mu}_{\mathrm{p}} e^{im\phi}= \rho_{\mathrm{n}}2\Omega \epsilon_{bzd} \delta v_\mathrm{n}^d .
\end{multline}

To simplify equation~(\ref{eom_e2}), we break the left-hand side into the following components. These simplifications rely on standard vector calculus identities and cylindrical coordinate basis properties:

\begin{enumerate}
    \item Divergence of the isotropic radial stress:
    \begin{equation}
    \label{part1}
    \nabla^a (g_{ab} \delta \tau_{rr} e^{im\phi}) = \delta \tau_{rr} \nabla^b e^{im\phi} + e^{im\phi} \frac{d \delta \tau_{rr}}{dr} \hat{r}^b,
    \end{equation}

    \item Divergence of the isotropic elastic term:
    \begin{multline}
     \nabla^a \left( g_{ab} e^{im\phi} 2\mu\left(\frac{\xi_r}{r} - \frac{d\xi_r}{dr}\right) \right) \\
    = 2\mu\left(\frac{\xi_r}{r} - \frac{d\xi_r}{dr}\right)\nabla^b e^{im\phi}
    + 2\mu e^{im\phi}\frac{d}{dr}\left(\frac{\xi_r}{r} - \frac{d\xi_r}{dr}\right)\hat{r}^b .
    \end{multline}

    \item Divergence of the anisotropic radial component:
    \begin{multline}
    \nabla^a \left(\hat{r}_a \hat{r}_b e^{im\phi} 2\mu\left(\frac{\xi_r}{r} - \frac{d\xi_r}{dr}\right)\right) = \frac{1}{r} e^{im\phi} 2\mu\left(\frac{\xi_r}{r} - \frac{d\xi_r}{dr}\right) \hat{r}^b \\
    + 2\mu e^{im\phi} \frac{d}{dr}\left(\frac{\xi_r}{r} - \frac{d\xi_r}{dr}\right) \hat{r}^b,
    \end{multline}

    using
    \begin{equation}
    \nabla^a \hat{r}_a = \frac{1}{r}, \quad \hat{r}^a \nabla_a \hat{r}_b = \frac{1}{r} \left( \delta_{ab} - \hat{z}_a \hat{z}_b - \hat{r}_a \hat{r}_b \right) \hat{r}^a.
    \end{equation}

    \item The vertical ($z$-axis) term vanishes:
    \begin{equation}
    \nabla^a (\hat{z}_a \hat{z}_b e^{im\phi} 2\mu \frac{\xi_r}{r}) = 0,
    \end{equation}

    \item The mixed radial-angular terms:
    \begin{multline}
    \nabla^a (r \hat{r}_a \nabla_b e^{im\phi} (\delta \tau'_{r\phi} + 2\mu \frac{\xi_\perp}{r})) = \nabla_b e^{im\phi} \delta \tau'_{r\phi} \\ + r \nabla_b e^{im\phi} \frac{d\delta \tau'_{r\phi}}{dr} + \nabla_b e^{im\phi} 2\mu \frac{d\xi_\perp}{dr},
    \end{multline}

    \begin{multline}
    \nabla^a (r \hat{r}_b \nabla_a e^{im\phi} (\delta \tau'_{r\phi} + 2\mu \frac{\xi_\perp}{r})) = \nabla_b e^{im\phi} \delta \tau'_{r\phi} + \nabla_b e^{im\phi} 2\mu \frac{\xi_\perp}{r} \\
    - m^2 \frac{\delta \tau'_{r\phi}}{r} e^{im\phi} \hat{r}_b - m^2 \frac{2\mu \xi_\perp}{r^2} e^{im\phi} \hat{r}_b,
    \end{multline}

    \item The Laplacian angular term:
    \begin{multline}
    \label{part7}
    \nabla^a (2\mu \xi_\perp r \nabla_a \nabla_b e^{im\phi}) = -2\mu \frac{d\xi_\perp}{dr} \nabla_b e^{im\phi} - 2\mu \frac{\xi_\perp}{r} \nabla_b e^{im\phi} \\
    - m^2 2\mu \frac{\xi_\perp}{r} \nabla_b e^{im\phi} + m^2 4\mu \frac{\xi_\perp}{r^2} e^{im\phi},
    \end{multline}

    using the identity:
    \begin{equation}
    \nabla_a \nabla^a \nabla_b e^{im\phi} = -m^2 \nabla_b \left(\frac{e^{im\phi}}{r^2}\right).
    \end{equation}
\end{enumerate}

Substituting equations~(\ref{part1})--(\ref{part7}) back into equation~(\ref{eom_e2}), and projecting along \(\hat{r}\) and \(\hat{\phi}\) directions, we obtain two components of the perturbed equation of motion:

\begin{multline}
\label{emo_elastic_r}
    \hat{r}: \left( \frac{d\delta \tau_{rr}}{dr} - \frac{2\mu}{r}\left(\frac{\xi_r}{r} - \frac{d\xi_r}{dr}\right) - \frac{m^2}{r} \delta \tau'_{r\phi} + m^2 2\mu \frac{\xi_\perp}{r^2}+ \delta \rho_{\mathrm{p}}\frac{d \Tilde{\mu_\mathrm{p}}}{dr}  \right) e^{im\phi} \\ = - \rho_{\mathrm{n}}2\Omega \tilde{v}(r) e^{im\phi},
\end{multline}

\begin{multline}
\label{eom_elastic_phi}
    \hat{\phi}: \left( \delta \tau_{rr} + 2\mu\left(\frac{\xi_r}{r} - \frac{d\xi_r}{dr}\right) + r \frac{d \delta \tau'_{r\phi}}{dr} + 2 \delta \tau'_{r\phi} - m^2 2\mu \frac{\xi_\perp}{r} \right) \nabla_\phi e^{im\phi} \\= -2i\rho_{\mathrm{n}}\Omega \tilde{u}(r) e^{im\phi}.
\end{multline}

Interestingly, all terms resulting from the radial dependence of $\mu$ cancel out.

These equations describe the dynamics of the perturbed elastic medium coupled to vortex motion. In our model, we consider only the $m=2$ perturbation. For $m=2$, equations~\eqref{emo_elastic_r} and ~\eqref{eom_elastic_phi} simplify to:

\begin{multline}
\label{emo_elastic_r_2}
    \hat{r}: \left( \frac{d\delta \tau_{rr}}{dr} - \frac{2\mu}{r}\left(\frac{\xi_r}{r} - \frac{d\xi_r}{dr}\right) - \frac{4}{r} \delta \tau'_{r\phi} + 8\mu \frac{\xi_\perp}{r^2} + \delta \rho_{\mathrm{p}}\frac{d \Tilde{\mu_\mathrm{p}}}{dr} \right) e^{i2\phi} \\= - \rho_{\mathrm{n}}2\Omega \tilde{v}(r) e^{i2\phi},
\end{multline}

\begin{multline}
\label{eom_elastic_phi_2}
    \hat{\phi}: \left( \delta \tau_{rr} + 2\mu\left(\frac{\xi_r}{r} - \frac{d\xi_r}{dr}\right) + r \frac{d \delta \tau'_{r\phi}}{dr} + 2 \delta \tau'_{r\phi} - 8\mu \frac{\xi_\perp}{r} \right) \nabla_\phi e^{i2\phi}\\ = -2i\rho_{\mathrm{n}}\Omega \tilde{u}(r)e^{i2\phi}.
\end{multline}

Taking the real part of equations~\eqref{emo_elastic_r_2} and \eqref{eom_elastic_phi_2}, we obtain the radial and azimuthal components of the equations of motion, as given in equations~\eqref{eq:emo_elastic_r_2_real_pomp} and \eqref{eq:eom_elastic_phi_2_real_pomp}.

%%%%%%%%%%%%%%%%%%%%%%%%%%%%%%%%%%%%%%%%%%%%%%%%%%

% Don't change these lines
\bsp	% typesetting comment
\label{lastpage}
\end{document}